\documentclass[journal]{IEEEtran}

\usepackage{amsmath}
\usepackage{graphicx}
\usepackage{amssymb}
\usepackage{epstopdf}
\usepackage{amsthm}
\usepackage{bm}
\usepackage{algorithm}
\usepackage{cite}
\usepackage[noend]{algpseudocode}
\usepackage{soul,color}
\usepackage{mathtools}
\usepackage{caption}
\usepackage{subcaption}
\usepackage{array}
\usepackage{enumitem}

\def\P{{\cal P}}

\def\J{{\cal J}}

\def\I{{\mathbb I}}

\def\L{{\cal L}}

\newcommand{\R}{\mathbb R}

\newcommand{\unorm}[1]{\lVert#1\rVert}
\newcommand{\uabs}[1]{\lvert#1\rvert}
\newcommand{\norm}[1]{\left\lVert#1\right\rVert}
\newcommand{\abs}[1]{\left\lvert#1\right\rvert}
\newcommand\numberthis{\addtocounter{equation}{1}\tag{\theequation}}

\newcommand{\topnew}{\top \hspace{-0.05cm}}
\DeclareMathOperator*{\argmin}{argmin}
\DeclareMathOperator*{\argmax}{argmax}
\DeclareMathOperator*{\vect}{vec}

\DeclareMathOperator*{\diag}{diag}

\DeclareMathOperator*{\corr}{corr}

\makeatletter
\newcommand\fs@betterruled{
  \def\@fs@cfont{\bfseries}\let\@fs@capt\floatc@ruled
  \def\@fs@pre{\vspace*{5pt}\hrule height.8pt depth0pt \kern2pt}
  \def\@fs@post{\kern2pt\hrule\relax}
  \def\@fs@mid{\kern2pt\hrule\kern2pt}
  \let\@fs@iftopcapt\iftrue}
\floatstyle{betterruled}
\restylefloat{algorithm}
\makeatother

\newcommand{\Scale}[2][4]{\scalebox{#1}{$#2$}}

\allowdisplaybreaks

\begin{document}

\title{Constrained Independent Vector Analysis with Reference for Multi-Subject fMRI Analysis}
\author{Trung Vu*, Francisco Laport*, Hanlu Yang, Vince D. Calhoun, and T\"{u}lay Adal{\i}
\thanks{This work is supported in part by the grants NIH R01MH118695, NIH R01MH123610, NIH R01AG073949, NSF 2112455, NSF 2316420, and Xunta de Galicia ED481B 2022/012.}
\thanks{Trung Vu, Francisco Laport, Hanlu Yang, and T\"{u}lay Adali are with Department of Computer Science and Electrical Engineering, University of Maryland, Baltimore County, MD 21250, USA (e-mail: \{trungvv, flopez2, hyang3, adali\}@umbc.edu). Francisco Laport is also with CITIC Research Center, University of A Coru\~{n}a, Campus de Elvi\~{n}a, 15071 A Coru\~{n}a, Spain.}
\thanks{Vince Calhoun is with the Tri-Institutional Center for Translational Research in Neuroimaging and Data Science (TReNDS), Georgia State University, Georgia Institute of Technology, and Emory University, Atlanta, GA 30303, USA (e-mail:vcalhoun@gsu.edu).}
\thanks{Trung Vu and Francisco Laport contributed equally to this work.}
}

\markboth{Preprint}%
{Vu \MakeLowercase{\textit{et al.}}: Constrained Independent Vector Analysis with Reference for Multi-Subject fMRI Analysis}

\maketitle
\begin{abstract}
Independent component analysis (ICA) is now a widely used solution for the analysis of multi-subject functional magnetic resonance imaging (fMRI) data. Independent vector analysis (IVA) generalizes ICA to multiple datasets, i.e., to multi-subject data, and in addition to higher-order statistical information in ICA, it leverages the statistical dependence across the datasets as an additional type of statistical diversity. As such, it preserves variability in the estimation of single-subject maps but its performance might suffer when the number of datasets increases. Constrained IVA is an effective way to bypass computational issues and improve the quality of separation by incorporating available prior information. Existing constrained IVA approaches often rely on user-defined threshold values to define the constraints. However, an improperly selected threshold can have a negative impact on the final results. This paper proposes two novel methods for constrained IVA: one using an adaptive-reverse scheme to select variable thresholds for the constraints and a second one based on a threshold-free formulation by leveraging the unique structure of IVA. We demonstrate that our solutions provide an attractive solution to multi-subject fMRI analysis both by simulations and through analysis of resting state fMRI data collected from 98 subjects --- the highest number of subjects ever used by IVA algorithms. Our results show that both proposed approaches obtain significantly better separation quality and model match while providing computationally efficient and highly reproducible solutions.
\end{abstract}
\begin{IEEEkeywords}
independent vector analysis, constrained IVA, multivariate Gaussian distribution, fMRI analysis.
\end{IEEEkeywords}

\section{Introduction}
\label{sec:intro}

\IEEEPARstart{I}{ndependent} component analysis (ICA) is a blind source separation (BSS) technique that decomposes a multivariate signal into statistically independent components. This data-driven approach has found fruitful applications in the analysis of neuroimaging data including functional magnetic resonance imaging (fMRI) \cite{mckeown1998analysis}, electroencephalography (EEG) \cite{onton2006imaging}, and magnetoencephalography (MEG) \cite{luckhoo2012inferring} data. In fMRI studies, ICA can extract either spatially or temporally independent components corresponding to a single subject \cite{calhoun2001spatial} while spatial ICA has been the dominant version \cite{calhoun2012multisubject}.

In generalization to analysis of multi-subject data, the group ICA method \cite{calhoun2001method} has been by far the most commonly used approach where the multi-subject fMRI data are temporally concatenated. Other approaches include spatial concatenation \cite{svensen2002ica} and tensor organization \cite{beckmann2005tensorial}. Once the group data is created, typically through two levels of dimension reduction using principal component analysis, ICA is applied to extract group-independent components, which can be then used to compute the subject-specific components by back-reconstruction, dual or other flavors of regression \cite{calhoun2001method, beckmann2009group, erhardt2011comparison}. 
The disadvantage of group ICA, however, is that it relies on the assumption of a common subspace among all subjects and hence, its ability to capture subject variability might be limited. 
Another powerful approach to multi-subject data analysis is independent vector analysis (IVA), which generalizes ICA to multi-subject fMRI analysis by exploiting the statistical dependencies across the subject datasets \cite{kim2006independent, anderson2011joint, adali2014diversity}. 
IVA has been shown to perform well in capturing subject variability \cite{michael2014preserving,laney2015capturing} and is competitive with ICA which has been extensively studied in this regard \cite{allen2012capturing}.
Additionally, as a by-product, dependent sources across the subjects are automatically aligned in IVA, avoiding the permutation ambiguity existing in ICA \cite{loesch2010robustness}.
Notwithstanding, one major drawback of IVA is that its performance degrades when the number of datasets increases or when the level of variability among the subjects is very low \cite{long2020independent, bhinge2019extraction}. 
Other approaches to group studies include multiset canonical correlation analysis (MCCA) \cite{kettenring1971canonical}, MultiView ICA \cite{richard2020modeling}, joint ICA (jICA) \cite{calhoun2006neuronal}, and tensor decomposition \cite{andersen2004structure,beckmann2005tensorial}.

ICA can be further improved in various ways by incorporating spatial constraints \cite{salman2019group}. Following a similar strategy, to improve the performance of IVA, constrained IVA has been developed as an effective way to incorporate prior knowledge (often about the sources or the mixing matrices) while also addressing the aforementioned limitations. Similar to constrained ICA \cite{lu2005approach, rodriguez2014general}, constrained IVA introduces (in)equality constraints to the cost function and utilizes the Lagrangian framework to solve the constrained optimization problem. A reliable set of constraints guides IVA algorithms to avoid sub-optimal solutions and increase the quality of source separation and of the estimated components by providing a better model match. There have been two major types of prior information (references): rough templates of the sources \cite{lu2005approach, zhang2008morphologically} or the mixing matrix \cite{calhoun2005semi, de2011spatially}. However, given the current emphasis on resting state fMRI data analyses, spatial constraints are more commonly used and are our focus here as well. If a reference is constructed properly, it is the one and only one that is closest to the desired signal in terms of a closeness measure. Therefore, when incorporated into the IVA framework, such references carry adequate information to distinguish the desired signal from artifacts and noise. 
One key challenge in constrained IVA is selecting a threshold parameter that enforces the closeness between the reference signal and the desired output. A feasible threshold depends on both the designed reference signal and the closeness measure. If the threshold is too small, the output may produce a sub-optimal solution. If the threshold is too large, the corresponding constraint might not be satisfied and cause the learning to become unpredictable. In the context of constrained ICA, Lu and Rajapakse \cite{lu2005approach} suggested using a small threshold initially and then gradually increasing the threshold. However, this method requires multiple runs of the algorithm, which is computationally expensive, especially for application to IVA. Recently, Bhinge~\textit{et~al.} \cite{bhinge2019extraction} studied constrained IVA with multivariate Laplace distributed sources (IVA-L-SOS) and proposed a heuristic scheme, named parameter-tuned constrained IVA (pt-cIVA), for selecting the constraint thresholds from a set of pre-defined values. This adaptive tuning scheme, while facilitating the automatic selection of threshold values, may lead to a sub-optimal solution where the threshold is much smaller than the desired value. 

In this paper, we present two attractive solutions to multi-subject fMRI analysis via constrained IVA with reference. First, we address the aforementioned issue with pt-cIVA by introducing an adaptive-reverse scheme for threshold selection, named adaptive-reverse constrained IVA (ar-cIVA). By alternating between a conservative scheme and an assertive scheme, our proposed approach allows the threshold values to increase when the constraints are easily satisfied and to decrease when the constraints are too difficult to achieve.
Furthermore, to eliminate the need for threshold selection, we propose a second method for constrained IVA that is threshold-free (tf-cIVA). The references are utilized as a regularization for the IVA cost function, in which not only the similarity between the reference and the corresponding source but also the similarity between that reference and the other sources are taken into account. 
In both methods, we leverage IVA with multivariate Gaussian sources to exploit second-order statistics (SOS) while utilizing the similarity between the sources and the references to account for higher-order statistics (HOS). Compared with IVA using a multivariate Laplace density model, both our approaches offer significantly faster runtime, with iteration complexity independent of the sample size.
We demonstrate the effectiveness of the proposed approaches through a number of experiments on both simulated fMRI-like data and real fMRI data with $98$ subjects. 
We emphasize that to the best of our knowledge, this is the highest number of subjects that have been used by the IVA framework when full correlation is taken into account as in our case.
The results show that our methods significantly outperform the unconstrained IVA method as well as existing constrained IVA methods. It is also promising that our approaches can be applied to large-scale data with a few hundred to thousand subjects, as well as other application domains in joint BSS. 
    
The rest of this paper is organized as follows. Section~\ref{sec:pre} provides a brief review of independent vector analysis, the multivariate Gaussian sources, and reference-constrained IVA. Then, Sections~\ref{sec:ar} and \ref{sec:tf} present our two proposed approaches to constrained IVA, namely adaptive-reverse constrained IVA and threshold-free constrained IVA, followed by the implementation details described in Section~\ref{sec:impl}. In Sections~\ref{sec:hybrid} and \ref{sec:fMRI}, we demonstrate the effectiveness of the proposed algorithms in simulated fMRI-like data and real fMRI data, respectively. Finally, Section~\ref{sec:conc} summarizes our work in this paper and discusses potential directions for future work.

\section{Preliminaries}
\label{sec:pre}

\noindent \textbf{Notation.}
Throughout the paper, we use the notations $\unorm{\cdot}_F$ and $\unorm{\cdot}_2$ to denote the Frobenius norm and the spectral norm of a matrix, respectively. Additionally, $\unorm{\cdot}$ is used on a vector to denote the Euclidean norm. Boldfaced symbols are reserved for vectors and matrices. The notation $(\cdot)^\topnew$ denotes the transpose of a matrix.
The $t \times t$ identity matrix is denoted by $\bm I_t$. The $t$-dimensional vector of all zeros and the $t$-dimensional vector of all ones are denoted by $\bm 0_t$ and $\bm 1_t$, respectively. In addition, the $i$th vector in the natural basis of $\R^t$ is denoted by $\bm e_i$.
The notation $\otimes$ denotes the Kronecker product between two matrices and $\vect(\cdot)$ denotes the vectorization of a matrix by stacking its columns on top of one another. Given an $n$-dimensional vector $\bm x$, $x_i$ denotes its $i$th element and $\diag(\bm x)$ denotes the $n \times n$ diagonal matrix with the corresponding diagonal entries $x_1,\ldots,x_n$. Similarly, for an $m \times n$ matrix $\bm X$, the $(i,j)$ entry of $\bm X$ is denoted by $X_{ij}$.

\subsection{Independent Vector Analysis (IVA)}

Consider $K$ datasets (subjects), each formed by $V$ samples (voxels) of linear mixtures of $N$ independent sources
\begin{align}
    \bm x^{[k]}(v) = \bm A^{[k]} \bm s^{[k]}(v) ,
\end{align}
for $k=1,\ldots,K$ and $v=1,\ldots,V$. Here, $\bm A^{[k]} \in \R^{N \times N}$ is an invertible mixing matrix for the $k$th dataset and $\bm s^{[k]}(v) = [s^{[k]}_1(v),\ldots,s^{[k]}_N(v)]^\topnew$ is the $v$th sample of the corresponding source vector. By stacking the $n$th source component across $K$ datasets, we introduce a key concept: the $n$th source component vector (SCV) as a $K$-dimensional random vector\footnote{For convenience, we use the same symbol $s_n^{[k]}$ to denote the random variable. The samples of $s_n^{[k]}$ are indicated by the index $v$ in $s_n^{[k]}(v)$.}
\begin{align*}
    \bm s_n = [s_n^{[1]},\ldots,s_n^{[K]}]^\topnew .
\end{align*}
An appropriate multivariate probability density function (pdf) of the SCV can take all order statistical information within and across the $K$ datasets into account.
The goal of IVA is to identify the independent SCVs via the estimation of $K$ demixing matrices of the form $\bm W^{[k]} = [\bm w_1^{[k]}, \ldots, \bm w_N^{[k]}]^\topnew \in \R^{N \times N}$.
Denote $\bm y^{[k]}(v)=\bm W^{[k]} \bm x^{[k]}(v)$ as the vector containing $N$ estimated sources for the $k$th dataset. The $n$th estimated SCV corresponding to the sample index $v$ is given by $\bm y_n(v) = [y^{[1]}_n(v),\ldots,y^{[K]}_n(v)]^\topnew \in \R^K$.
Assuming the samples are independently and identically distributed (iid) and using the maximum likelihood principle \cite{adali2014diversity}, one can write the IVA cost as minimizing the negative log-likelihood w.r.t. $\bm W = \{ \bm W^{[k]} \}_{k=1}^K$ and $\bm \Sigma = \{ \bm \Sigma_n \}_{n=1}^N$ 
\begin{align*}
    \Scale[.93]{\J_{\text{IVA}} (\bm W, \bm \Sigma) \triangleq -\frac{1}{V} \displaystyle{\sum_{v,n}} \log p_n(\bm y_n(v) \mid \bm \Sigma_n) - \displaystyle{\sum_{k}} \log \uabs{\det{\bm W^{[k]}}}} , \numberthis \label{equ:IVA_cost}
\end{align*}
where $p_n$ denotes the pdf of the $n$th SCV. In this formulation, there are no restrictions on the $\bm W^{[k]}$ beyond being invertible. 

\begin{table*}[t]
\caption{Formulas of the IVA cost function, the augmented Lagrange function, and its gradient.}
\label{tbl:formulas}
\centering
\begin{minipage}{\textwidth}
\begin{align*}
    &\J_{\text{IVA}} (\bm W, \bm \Sigma) = \frac{NK}{2} \log(2\pi) + \frac{1}{2} \sum_{n=1}^N \log \abs{\det(\bm \Sigma_n)} + \frac{1}{2} \sum_{n=1}^N \sum_{k,l=1}^K (\bm e_k^\topnew \bm \Sigma_n^{-1} \bm e_l) (\bm w_n^{[k]})^\topnew \bigl(\frac{1}{V} \bm X^{[k]} (\bm X^{[l]})^\topnew \bigr) \bm w_n^{[l]} - \sum_{k=1}^K \log \uabs{\det{(\bm W^{[k]})}} \numberthis \label{equ:IVAG_cost} \\
    &\L_{\gamma, \bm \rho} (\bm W, \bm \Sigma, \bm \mu) = \J_{\text{IVA}} (\bm W, \bm \Sigma) + \frac{1}{2\gamma} \sum_{n=1}^M \sum_{k=1}^K \Biggl( \biggl( \max \Bigl( 0, \mu_{n}^{[k]} + \gamma \bigl( \rho_n^{[k]} - \epsilon(\bm r_n,\bm y_n^{[k]}) \bigr) \Bigr) \biggr)^2 - (\mu_{n}^{[k]})^2 \Biggr) \numberthis \label{equ:L_augmented_IVA2} \\
    &\frac{\partial \L_{\gamma, \bm \rho}}{\partial \bm w_n^{[k]}} = \sum_{l=1}^K \bigl( \frac{1}{V} \bm X^{[k]} (\bm X^{[l]})^\topnew \bigr) \bm w_n^{[l]} \bm e_l^\topnew \bm \Sigma_n^{-1} \bm e_k - \frac{\bm d_n^{[k]}}{(\bm d_n^{[k]})^\topnew \bm w_n^{[k]}} - \I_{n\leq M} \max \Bigl( 0, \mu_{n}^{[k]} + \gamma \bigl( \rho_n^{[k]} - \epsilon(\bm r_n,\bm y_n^{[k]}) \bigr) \Bigr) \bm X^{[k]} \frac{\partial \epsilon(\bm r_n,\bm y_n^{[k]})}{\partial \bm y_n^{[k]}} \numberthis \label{equ:grad_J_w_constrained}
\end{align*}
\end{minipage}
\end{table*}

The multivariate Gaussian distribution (MGD) provides an attractive solution to model SCV pdfs in terms of complexity and allows taking full SOS into account. Assuming each estimated SCV $\bm y_n$ follows an MGD with zero mean and  covariance matrix $\bm \Sigma_n \in \R^{K \times K}$, the IVA cost in (\ref{equ:IVA_cost}) can be rewritten as (\ref{equ:IVAG_cost}) in Table~\ref{tbl:formulas}, where $\bm X^{[k]} = [\bm x^{[k]}(1),\ldots,\bm x^{[k]}(V)]$ is the $N \times V$ data matrix.
In \cite{anderson2011joint}, Anderson~\textit{et~al.} study the theoretical properties (e.g., local stability and identifiability conditions) of the IVA framework with MGD source model, i.e., IVA-G, and demonstrate its effectiveness in joint BSS. 

In application to fMRI analysis, since underlying sources are more likely to be super-Gaussian \cite{adali2014diversity,long2020independent}, the use of only SOS might come across as a limitation. For example, the multivariate Laplace distribution (MLD) has been shown to provide a better model match to fMRI sources \cite{calhoun2012multisubject,bhinge2019extraction}.
However, this approach is computationally expensive since its iteration complexity depends on the number of data samples. In multi-subject fMRI data analysis, IVA with MLD methods such as IVA-L-SOS and its constrained variants have only been applied to medium-scale settings of no more than $64$ subjects and $20$ components \cite{long2021relationship}. In this work, we use a larger dataset of $98$ subjects and select an order of $60$ components. We demonstrate how to guide the estimation by introducing reference signals and how the model match is maintained while still achieving computational efficiency. In subsequent sections, we will simply refer to the IVA-G cost in (\ref{equ:IVAG_cost}) as IVA cost for convenience.



\subsection{Constrained IVA with Reference}
\label{sec:cIVA}

In constrained IVA, we consider a set of reference signals $\{ \bm r_n \}_{n=1}^M \subset \R^V (M \leq N)$ that can be used as prior constraints to guide the separation of sources. 
For the $k$th dataset, the $n$th estimated source is given by $\bm y_n^{[k]} = [y_n^{[k]}(1),\ldots,y_n^{[k]}(V) ]^\topnew$. The idea here is to ensure that $\bm r_n$ has a higher correlation with its corresponding SCV than any other SCVs in the same dataset, i.e.,
\begin{align} \label{equ:eps_mn}
    \epsilon(\bm r_n,\bm y_n^{[k]}) > \epsilon(\bm r_n,\bm y_m^{[k]}) \quad \forall m \neq n ,
\end{align}
where $\epsilon : \R^V \times \R^V \to [0,1]$ is some similarity measure, $n=1,\ldots,M$, and $m=1,\ldots,N$. As an example, $\epsilon(\cdot)$ can be chosen as the absolute value of Pearson correlation
\begin{align} \label{equ:Pearson}
    \epsilon(\bm a, \bm b) = \abs{\corr(\bm a, \bm b)} = \frac{\abs{\bm a^\topnew \bm b}}{\norm{\bm a} \norm{\bm b}} .
\end{align}
A common approach to implementing such constraints is via a pre-defined threshold parameter $\rho$ \cite{lu2005approach,rodriguez2014general,bhinge2017non}. By selecting an appropriate value of $\rho$ such that
\begin{align} \label{equ:rho_middle}
    \epsilon(\bm r_n,\bm y_n^{[k]}) \geq \rho > \epsilon(\bm r_n,\bm y_m^{[k]}) \quad \forall m \neq n ,
\end{align}
only one independent component is extracted as the closest one to the reference signal. Thus, the thresholding-constrained formulation is proposed in \cite{bhinge2017non} as
\begin{align*} 
    \min_{\bm W, \bm \Sigma} \J_{\text{IVA}} (\bm W, \bm \Sigma) \quad \text{s.t. } \epsilon(\bm r_n,\bm y_n^{[k]}) \geq \rho_n \quad \forall n, \numberthis \label{equ:cIVA}
\end{align*} 
where $n=1,\ldots,M$ and $k=1,\ldots,K$.
The major disadvantage of formulation (\ref{equ:cIVA}) is that the best values for the threshold parameters are often unknown in practice. If $\rho_n$ is too small, the output may produce a different component. If $\rho_n$ is too large, the estimate might not yield a desired component because the corresponding constraint causes the learning to become unpredictable. 
Hence, an ideal value of $\rho_n$ is the one that is closest to the similarity between the reference $\bm r_n$ and the true source $\bm s_n^{[k]}$, i.e., $\epsilon(\bm r_n, \bm s_n^{[k]})$.
To address this issue, an adaptive scheme to select $\rho$, pt-cIVA, has been proposed in \cite{bhinge2019extraction}. The idea is to use a set of predefined thresholds $\P$ and at each iteration, pick a value that is closest to the similarity value between the reference $\bm r_n$ and the estimated sources $\bm y_n^{[k]}$:
\begin{align} \label{equ:rho_pt}
    \rho_n = \argmin_{ \rho \in \P } \min_{1\leq k \leq K} \abs{\rho-\epsilon(\bm r_n, \bm y_n^{[k]})} .
\end{align}
While this heuristic was shown to improve the performance of constrained IVA \cite{bhinge2019extraction}, it may lead to a sub-optimal solution where $\rho_n$ can be much smaller than $\epsilon(\bm r_n, \bm s_n^{[k]})$. 
Indeed, if at some iteration, the threshold is always selected such that it is smaller than or equal to $\epsilon(\bm r_n, \bm y_n^{[k]})$, the constraints in (\ref{equ:cIVA}) will be automatically satisfied and will have no effect on increasing $\epsilon(\bm r_n, \bm y_n^{[k]})$ in the next iteration.
Another issue with formulation (\ref{equ:cIVA}) is that the threshold $\rho_n$ does not depend on $k$ and hence, does not consider the case where $\epsilon(\bm r_n, \bm y_n^{[k]})$ has a different threshold from $\epsilon(\bm r_n, \bm y_n^{[l]})$, for $k \neq l$. To accommodate the variability among the subjects within an SCV, one needs to impose different thresholds for different levels of closeness $\epsilon(\cdot)$ in the constraints.

\section{Adaptive-Reverse Constrained IVA}
\label{sec:ar}

We introduce an adaptive-reverse scheme for selecting the constraint thresholds that significantly improves the performance of pt-cIVA while maintaining the same computational complexity per iteration.
First, we extend (\ref{equ:cIVA}) to a more flexible constrained formulation that takes into account the subject variability across components 
\begin{align*} 
    \min_{\bm W, \bm \Sigma} \J_{\text{IVA}} (\bm W, \bm \Sigma) \quad \text{s.t. } \epsilon(\bm r_n,\bm y_n^{[k]}) \geq \rho_n^{[k]} \quad \forall n,k . \numberthis \label{equ:cIVA_k}
\end{align*}
It is emphasized that (\ref{equ:cIVA}) uses the same threshold $\rho_n$ for all $K$ subjects in the $n$th component while (\ref{equ:cIVA_k}) uses different thresholds $\rho_n^{[k]}$ for each subject in the $n$th component.
Second, we propose an adaptive scheme to select $\rho_n^{[k]}$ that alternates between two principles: 
(i) choosing the smallest value that does not satisfy the constraint
\begin{align} \label{equ:rho_min}
    \rho_n^{[k]} = \argmin \{ \rho \in \P \mid \rho > \epsilon(\bm r_n, \bm y_n^{[k]}) \} , 
\end{align}
and (ii) choosing the largest value that satisfies the constraint  
\begin{align} \label{equ:rho_max}
    \rho_n^{[k]} = \argmax \{ \rho \in \P \mid \rho \leq \epsilon(\bm r_n, \bm y_n^{[k]}) \} .
\end{align}
On the one hand, (\ref{equ:rho_min}) creates an over-tight constraint that forces the value of $\epsilon(\bm r_n, \bm y_n^{[k]})$ to increase after each iteration. On the other hand, (\ref{equ:rho_max}) creates a feasible problem where each constraint is always satisfied. Using the appropriate principle at each iteration, the desired value of the threshold --- that is close to $\epsilon(\bm r_n, \bm s_n^{[k]})$ --- can be recovered. 

\begin{algorithm}[t]
\caption{Adaptive-Reverse Constrained IVA (ar-cIVA)}
\label{algo:vec_grad_algo}
\begin{algorithmic}[1]
\Require{$\{\bm X^{[k]}\}_{k=1}^K \subset \R^{N \times V}, \{ \bm r_n \}_{n=1}^M \subset \R^V$, $\gamma$, $\mu_{\max}$}
\Ensure{$\bm W, \bm \Sigma$}

\State Set the current scheme to (\ref{equ:rho_min})
\Repeat
\For{$n=1,\ldots,N$}
\For{$k=1,\ldots,K$}
\State Compute $\hat{\bm \Sigma}_n^{-1}$ 
\State Update $\mu_{n}^{[k]}$ using (\ref{equ:aug_update_alpha}) and (\ref{equ:aug_update_mu})
\If{$\mu_{n}^{[k]} \geq \mu_{\max}$}
\State Switch the current scheme to  (\ref{equ:rho_max})
\ElsIf{$\mu_{n}^{[k]} \leq 0$}
\State Switch the current scheme to (\ref{equ:rho_min})
\Else 
\State Keep the current scheme
\EndIf
\State Select $\rho_{n}^{[k]}$ based on the current scheme
\State Compute $d \bm w_n^{[k]} = {\partial \L_{\gamma, \bm \rho}}/{\partial \bm w_n^{[k]}}$ using (\ref{equ:grad_J_w_constrained})
\State Project $\tilde{d} \bm w_n^{[k]} = (\bm I_n - \bm w_n^{[k]} (\bm w_n^{[k]})^\topnew) d \bm w_n^{[k]}$
\State Update $\bm w_n^{[k]} = \bm w_n^{[k]} - \eta \frac{\tilde{d} \bm w_n^{[k]}}{\unorm{\tilde{d} \bm w_n^{[k]}}}$
\State Normalize $\bm w_n^{[k]} = \frac{\bm w_n^{[k]}}{\unorm{\bm w_n^{[k]}}}$
\State Update $[\hat{\bm \Sigma}_n]_{kl}$ for $l=1,\ldots,K$
\EndFor
\EndFor
\Until{convergence}
\end{algorithmic}
\end{algorithm}

\subsection{Augmented Lagrangian Method with Decoupling}

To solve (\ref{equ:cIVA_k}) as an inequality-constrained optimization, we utilize the augmented Lagrangian method and a decoupling method that enables sequential updates of each row of individual demixing matrices.
The augmented Lagrangian function is given in (\ref{equ:L_augmented_IVA2}), where $\bm \mu \in \R^{M \times K}$ is the Lagrange multiplier and $\gamma>0$ is the scalar penalty parameter as in \cite{lu2000constrained} where the framework is used for ICA.
It can be shown \cite{bertsekas2014constrained} that for sufficiently large $\gamma$, the solution of (\ref{equ:L_augmented_IVA2}) coincides with the solution of (\ref{equ:cIVA_k}).
At the $i$th iteration, we update the parameters to minimize $\L_{\gamma, \bm \rho}$ based on their current values as follows
\begin{align*} 
    &(\bm W^{i+1}, \bm \Sigma^{i+1}) = \textstyle{\argmin_{\bm W, \bm \Sigma}} \L_{\gamma, \bm \rho} \bigl(\bm W, \bm \Sigma , \bm \mu^i \bigr) , \numberthis \label{equ:aug_update} \\
    &(\alpha_{n}^{[k]})^{i+1} = (\mu_{n}^{[k]})^i + \gamma \bigl((\rho_n^{[k]})^i - \epsilon(\bm r_n, (\bm y_n^{[k]})^{i+1})\bigr) , \numberthis \label{equ:aug_update_alpha} \\
    &(\mu_{n}^{[k]})^{i+1} = \max \bigl(0,  (\alpha_{n}^{[k]})^{i+1} \bigr) , \numberthis \label{equ:aug_update_mu}
\end{align*}
where $\mu_{n}^{[k]}$ is the $(n,k)$-entry of $\bm \mu$.
In (\ref{equ:aug_update}), the value of $\bm \Sigma_n$ that minimizes $\L_{\gamma, \bm \rho}(\cdot)$ is given by
\begin{align} \label{equ:argmin_Sigma}
    \hat{\bm \Sigma}_n = \frac{1}{V} \sum_{v=1}^V \bm y_n(v) \bm y_n(v)^\topnew = \argmin_{\bm \Sigma_n} \J_{\text{IVA}} (\bm W, \bm \Sigma) . 
\end{align}
Additionally, to update $\bm W$, we utilize the vector gradient method IVA-G-V in \cite{anderson2011joint} and derive the gradient of the augmented Lagrange function as follows. First, we rewrite the term $\log \abs{\det{\bm W^{[k]}}}$ in (\ref{equ:IVAG_cost}) as the sum of two terms $\log \uabs{(\bm d_n^{[k]})^\topnew \bm w_n^{[k]}} + \log (\det ( \tilde{\bm W}_n^{[k]} (\tilde{\bm W}_n^{[k]})^\topnew ))/2$ where $\tilde{\bm W}_n^{[k]}$ is the $(N-1)\times N$ matrix obtained by removing the $n$th row from $\bm W^{[k]}$ and $\bm d_n^{[k]} \in \R^N$ satisfies $\tilde{\bm W}_n^{[k]} \bm d_n^{[k]} = \bm 0_{N-1}$.
This technique \cite{li2007nonorthogonal} is often referred to as the \textit{decoupling trick}, enabling the derivation of the gradient of $\J_{\text{IVA}}$ w.r.t. $\bm w_n^{[k]}$
\begin{align*} 
    \frac{\partial \J_{\text{IVA}}}{\partial \bm w_n^{[k]}} &= \frac{1}{V} \sum_{v=1}^V \bm x^{[k]}(v) (\bm y_n(v))^\topnew \bm \Sigma_n^{-1} \bm e_k - \frac{\bm d_n^{[k]}}{(\bm d_n^{[k]})^\topnew \bm w_n^{[k]}} .
\end{align*}
The advantage of this decoupling procedure is that one can avoid the dependence on the number of samples by recognizing that $y_n^{[k]}(v) = (\bm w_n^{[k]})^\topnew \bm x_n^{[k]}$ and pre-computing the sample covariance matrix $\hat{\bm R}_x^{kl} = \frac{1}{V-1} \bm X^{[k]} (\bm X^{[l]})^\topnew = \frac{1}{V-1} \sum_t \bm x^{[k]}(v) (\bm x^{[l]}(v))^\topnew$.
Second, summing the gradient of the cost function and the gradient of the constraint, we obtain the gradient of the augmented Lagrange function $\L_{\gamma, \bm \rho}$ w.r.t to $\bm w_n^{[k]}$ in (\ref{equ:grad_J_w_constrained}), for $n=1,\ldots,N$ and $k=1,\ldots,K$. In this formula, $\I_{n\leq M}$ is the indicator of the event $n \leq M$. 
When the Pearson correlation is used as the similarity measure, the last term in (\ref{equ:grad_J_w_constrained}) can be further simplified and can be computed independent of the sample size $V$.
Finally, certain refinements as suggested in \cite{li2010independent} (e.g., projecting the gradient onto the tangent space to the unit sphere, normalizing the gradient norm, and projecting the demixing vector back onto the unit sphere) are incorporated into the algorithm.

\subsection{Adaptive-Reverse Scheme for Constraint Thresholds}

We can now define the adaptive-reverse scheme for selecting the values of $\rho_n^{[k]}$.
In (\ref{equ:aug_update_alpha}) and (\ref{equ:aug_update_mu}), we note that the value of $\mu_n^{[k]}$ increases when the constraint $\epsilon(\bm r_n,\bm y_n^{[k]}) \geq \rho_n^{[k]}$ is violated and decreases when the constraint holds. Therefore, using the argmin scheme (\ref{equ:rho_min}) will increase the value of the Lagrange multipliers toward $+\infty$ while using the argmax scheme (\ref{equ:rho_max}) will decrease their values toward $0$. To combine the advantage of both approaches, we propose an adaptive-reverse scheme that determines the scheme based on the values of the Lagrange multipliers. In particular, when $\mu_n^{[k]}$ exceeds a certain value $\mu_{\max}$, we switch from the argmin scheme to the argmax scheme. Conversely, when $\mu_n^{[k]}$ goes down to $0$, we switch from the argmax scheme to the argmin scheme. We summarize the adaptive-reverse for constrained IVA (ar-cIVA) in Algorithm~\ref{algo:vec_grad_algo}.

\section{Threshold-Free Constrained IVA}
\label{sec:tf}

In this section, we propose a novel formulation of constrained IVA that eliminates the need for threshold parameters. Our idea is to maximize the similarity between the reference $\bm r_n$ and the corresponding estimated source component (corresponding-component similarity), and at the same time, promote the dissimilarity between that reference $\bm r_n$ and the other estimated component $\bm y_m^{[k]}$ (cross-component similarity), for all $m \neq n$. Thus, we introduce a regularization term 
\begin{align*}
    \J_{\text{ref}}(\bm W) = \sum_{n=1}^M \sum_{k=1}^K \Biggl( \sum_{\substack{m=1 \\ m \neq n}}^M \epsilon^2 (\bm r_n,\bm y_m^{[k]}) - \epsilon^2 (\bm r_n,\bm y_n^{[k]}) \Biggr) . \numberthis \label{equ:J_ref}
\end{align*}
The new objective function is hence a linear sum of the IVA cost function and the regularization
\begin{align} \label{equ:L_lambda}
    \L_{\lambda} (\bm W, \bm \Sigma) = \J_{\text{IVA}} (\bm W, \bm \Sigma) + \frac{\lambda}{2} \J_{\text{ref}}(\bm W) ,
\end{align}
where $\lambda>0$ is the regularization parameter. By selecting an appropriate value for $\lambda$ (via parameter tuning), we can balance the trade-off between the IVA cost (minimizing the correlation between the source components) and the regularization term (maximizing the correlation between the components and the reference signals).
Our formulation in (\ref{equ:L_lambda}) is similar to the multi-objective function optimization framework in \cite{du2013group}. In their work, Du and Fan introduced an improved version of constrained ICA by optimizing two conflicting cost functions: one that maximizes the independence among the components and one that maximizes the closeness between the components and their corresponding references. Nonetheless, compared with the approach in \cite{du2013group}, our proposed method not only generalizes constrained ICA to constrained IVA but also introduces the cross-component similarity to the objective function. This promotes the solution in which there is one and only one independent component that is closest to each reference.

From (\ref{equ:J_ref}), the gradient of $\J_{\text{ref}}(\cdot)$ w.r.t. $\partial \bm w_n^{[k]}$, for $n=1,\ldots,N$ and $k=1,\ldots,K$, is given by
\begin{align*}
    \frac{\partial \J_{\text{ref}}}{\partial \bm w_n^{[k]}} = 2 \mathbb{I}_{n\leq M} \biggl( \sum_{\substack{m=1 \\ m \neq n}}^M &\epsilon(\bm r_m,\bm y_n^{[k]}) \frac{\partial \epsilon(\bm r_m,\bm y_n^{[k]})}{\partial \bm w_n^{[k]}} \\
    &- \epsilon(\bm r_n,\bm y_n^{[k]}) \frac{\partial \epsilon(\bm r_n,\bm y_n^{[k]})}{\partial \bm w_n^{[k]}} \biggr) . \numberthis \label{equ:grad_J_ref}
\end{align*}
Thus, the gradient of $\L_{\lambda}(\cdot)$ is the sum of ${\partial \J_{\text{IVA}}}/{\partial \bm w_n^{[k]}}$ and $\lambda {\partial \J_{\text{ref}}}/{\partial \bm w_n^{[k]}}$.
The vector-gradient method to minimize $\L_{\lambda}(\cdot)$, named tf-cIVA, is described in Algorithm~\ref{algo:lambda_cIVA}. 
Compared with the thresholded formulation for constrained IVA in (\ref{equ:cIVA_k}), the regularized formulation does not require threshold parameters as well as other hyperparameters for the augmented Lagrange method (i.e., $\gamma$ and $\mu_{\max}$). 

\begin{algorithm}[t]
\caption{Threshold-Free Constrained IVA (tf-cIVA)}
\label{algo:lambda_cIVA}
\begin{algorithmic}[1]
\Require{$\{\bm X^{[k]}\}_{k=1}^K \subset \R^{N \times V}, \{ \bm r_n \}_{n=1}^M \subset \R^V$, $\lambda$}
\Ensure{$\bm W, \bm \Sigma$}

\Repeat
\For{$n=1,\ldots,N$}
\For{$k=1,\ldots,K$}
\State Compute $\hat{\bm \Sigma}_n^{-1}$
\State Compute $d \bm w_n^{[k]} = {\partial \L_{\lambda}}/{\partial \bm w_n^{[k]}}$ based on (\ref{equ:grad_J_ref})
\State Update $\bm w_n^{[k]}$ using $d \bm w_n^{[k]}$
\State Update $[\hat{\bm \Sigma}_n]_{kl}$ for $l=1,\ldots,K$
\EndFor
\EndFor
\Until{convergence}
\end{algorithmic}
\end{algorithm}

\section{Implementation and Evaluation}
\label{sec:impl}

\noindent \textbf{Compared Methods.}
We compare our new algorithms against the following methods: IVA-G-V \cite{anderson2011joint} for unconstrained IVA, cIVA-fixed for constrained IVA with fixed threshold \cite{bhinge2017non}, and its adaptive thresholding version (pt-cIVA) \cite{bhinge2019extraction}. The IVA-G-V algorithm for the unconstrained problem can be viewed as a baseline where no prior knowledge about the sources is used. For pt-cIVA, we note that the proposed version in \cite{bhinge2019extraction} uses MLD for the SCVs, which is significantly slower than MGD. Indeed, the IVA-L-SOS versions do not finish within $2$ weeks while the IVA-G versions run for a few hours, using the same setting in our simulation. Therefore, we reimplement pt-cIVA with MGD to make its computational time comparable with other methods. In addition, we use the set of pre-defined thresholds $\P_{pt} = \{ 0.001, 0.1, 0.2, \ldots, 0.9 \}$ and the penalty parameter $\gamma=3$ as specified by the authors in \cite{bhinge2019extraction}. For ar-cIVA, we use a finer set of pre-defined thresholds $\P_{ar} = \{0.01, 0.02, \ldots, 0.99\}$.\footnote{We also tried the finer set of thresholds for pt-cIVA but there was no significant difference. Hence, we present the results with default options for pt-cIVA in this work.} Furthermore, we set the penalty parameter $\gamma=100$ and the cut-off value for the Lagrange multiplier $\mu_{\max}=1$. For tf-cIVA, we use the tuned values for the regularization parameter: $\lambda=1$ for the simulated data and $\lambda=100$ for the real data.
All algorithms use the same initial step size $\eta$ for the gradient updates and a decay scheme that decreases $\eta$ by a factor of $0.95$ when the objective function does not decrease at a certain iteration. In addition, the stopping criteria for all algorithms are based on the change in $\bm W$ at each iteration \cite{anderson2011joint}
\begin{align*}
    \max_{k,n} \Bigl\{ 1 - \abs{{(\bm w_n^{[k]})^i}^\topnew (\bm w_n^{[k]})^{i+1}} \Bigr\} < \epsilon ,
\end{align*}
where $\epsilon=10^{-6}$ throughout this work.

\noindent \textbf{Evaluation metric.}
To evaluate the performance of different IVA algorithms, we use the following metrics: 

\noindent - \underline{Joint inter-symbol-interference} (joint-ISI) is introduced in \cite{anderson2011joint}, which is an extension of the normalized inter-symbol-interference (ISI) in the context of ICA \cite{amari1995new}. Let $\bm G^{[k]} = \bm W^{[k]} \bm A^{[k]}$, for $k=1,\ldots,K$ be the global demixing-mixing matrices and $\uabs{\bm G^{[k]}}$ be the absolute matrix with the $(m,n)$ entry being $\uabs{G^{[k]}_{mn}}$. The joint-ISI is defined as the ISI of the mean absolute value matrix $\bm G = 1/K \sum_{k=1}^K \uabs{\bm G^{[k]}}$, i.e., $\text{joint-ISI} \bigl( \bm G^{[1]}, \ldots, \bm G^{[K]} \bigr) = \text{ISI} (\bm G)$, where 
\begin{align*}
    \Scale[.95]{\text{ISI} (\bm G) = \frac{\displaystyle{\sum_{i=1}^N} \Bigl( \frac{\sum_{j=1}^N \abs{G_{ij}}}{\max_p \abs{G_{ip}}} - 1 \Bigr) + \sum_{j=1}^N \Bigl( \frac{\sum_{i=1}^N \abs{G_{ij}}}{\max_p \abs{G_{pj}}} - 1 \Bigr)}{2N(N-1)} .}
\end{align*}
When the sources are jointly separated for all datasets, the estimated global matrices for all datasets should be close to an identity matrix up to the same permutation. Thus, the joint-ISI closer to $0$ indicates better performance.

\noindent - \underline{Cross joint inter-symbol-interference} (cross-joint-ISI) measures the consistency of the components across $R$ runs. Let $\bm W_r^{[k]}$ be the $k$th demixing matrix of the $r$th run. In \cite{long2018consistent}, the cross-joint-ISI of the $i$th run and the $j$th run is defined as
\begin{align*}
    \Scale[.97]{\text{cross-joint-ISI}_{ij} (\{ \bm W_r^{[k]} \}_{r=1,k=1}^{R,K}) = \text{joint-ISI}(\bm P_{ij}^{[1]},\ldots,\bm P_{ij}^{[K]}) ,}
\end{align*}
where $\bm P_{ij}^{[k]} = \bm A_i^{[k]} \bm W_j^{[k]}$ and $\bm A_i^{[k]} = (\bm W_i^{[k]})^{-1}$. The cross-joint-ISI of the $i$th run is computed by averaging all its pairwise cross-joint-ISI values
\begin{align*}
    \text{cross-joint-ISI}_{i} = \frac{1}{R} \sum_{j=1, j\neq i}^R \text{cross-joint-ISI}_{ij} .
\end{align*}
Note that cross-joint-ISI can be computed when there is no ground truth available as it only depends on the demixing matrices. On the other hand, joint-ISI requires the true demixing matrices in its evaluation.

\noindent - \underline{Similarity factor} (SF) measures the average of the squares of the correlation between the estimated source and the corresponding ground truth:
\begin{align*}
    SF = \Bigl( \frac{1}{MK}\sum_{n=1}^M \sum_{k=1}^K \bigl( \epsilon (\bm s_n^{[k]}, \bm y_n^{[k]}) \bigr)^2 \Bigr)^{1/2} .
\end{align*}
The index $n$ runs from $1$ to $M$, meaning that only source components with corresponding reference signals are used.
A lower value of this metric indicates poor estimation of the sources as well as poor source alignment across the datasets.

\section{Hybrid Simulation Results}
\label{sec:hybrid}

This section compares the performance of the two proposed methods with the three aforementioned IVA algorithms using simulated fMRI-like data by changing the number of subjects and reference signals. Our goal is to better understand the behavior of these algorithms in different types of fMRI datasets. In the next section, we demonstrate an application with a practical fMRI dataset. 


\begin{figure}
    \centering
    \includegraphics[width=.8\columnwidth]{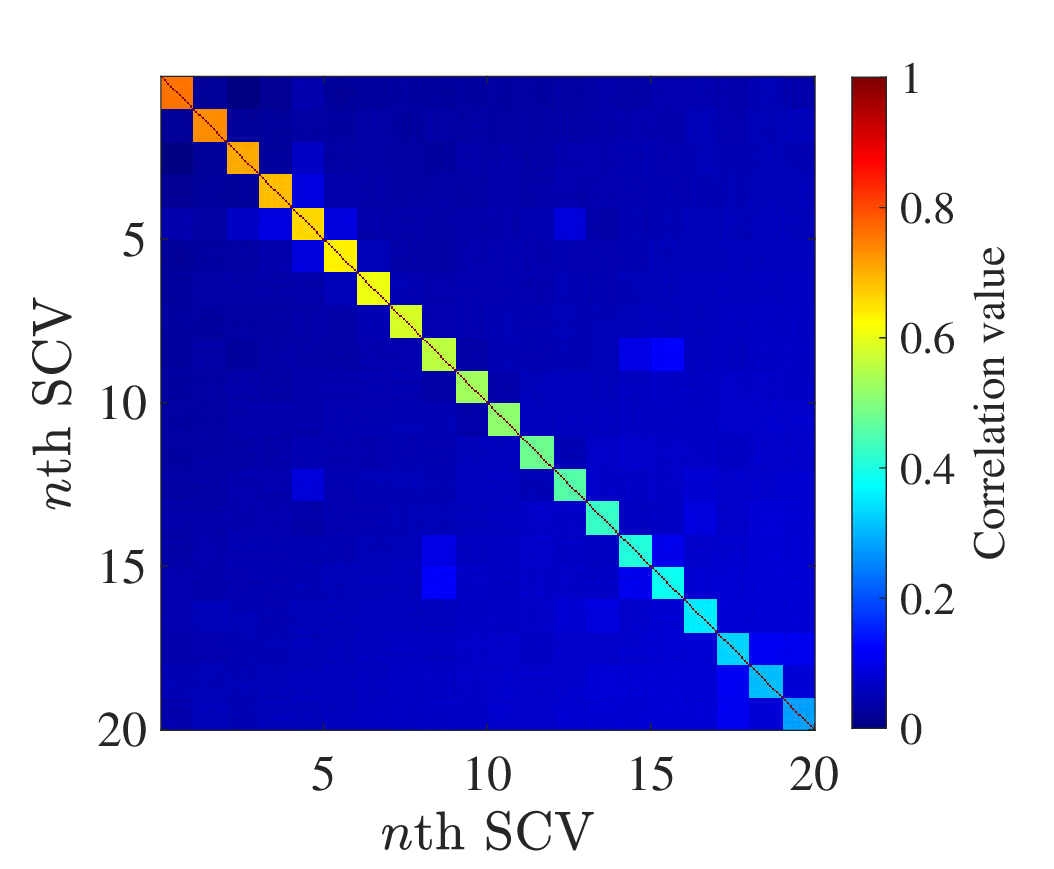}
    \caption{A visualization of the full SCV covariance matrix using model (\ref{equ:s_hybrid}) with $N=20, K=20$, $V=58515$, $\mu_0=0.1$, $\mu_1=0.2$, and $\bm \varphi=[\varphi_1,\varphi_2,\ldots,\varphi_N]^\topnew$ as a linearly spaced vector in the range $[0.3,0.9]$. From (\ref{equ:s_hybrid}), one can compute the values of the $20$ diagonal blocks in this matrix, ranging from $0.28$ (bottom-right) to $0.76$ (top-left). 
    }
    \label{fig:hybrid_data}
\end{figure}

\begin{figure*}[t]
    \centering
    \begin{subfigure}[b]{0.24\textwidth}
        \centering
        \includegraphics[width=\textwidth]{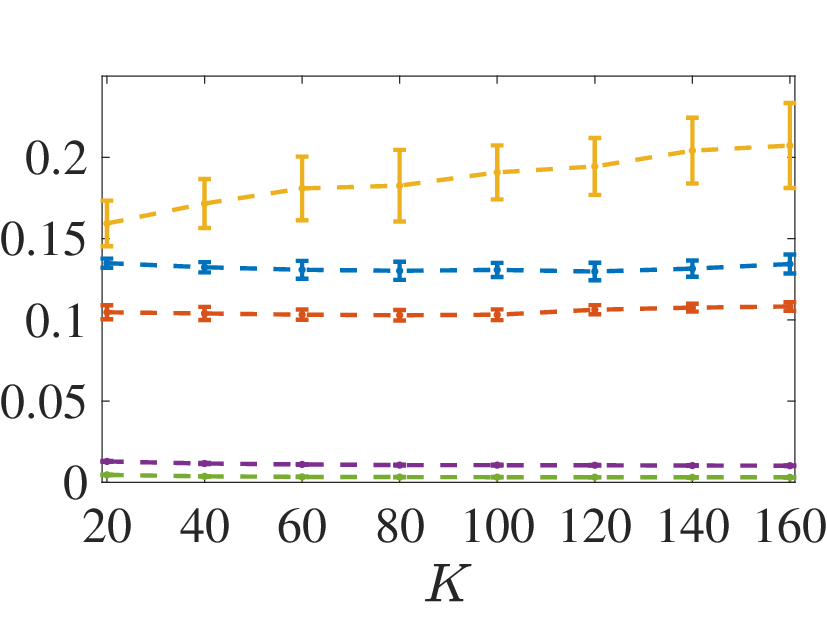}
        \caption{Joint-ISI}
    \end{subfigure}
    \begin{subfigure}[b]{0.24\textwidth}
        \centering
        \includegraphics[width=\textwidth]{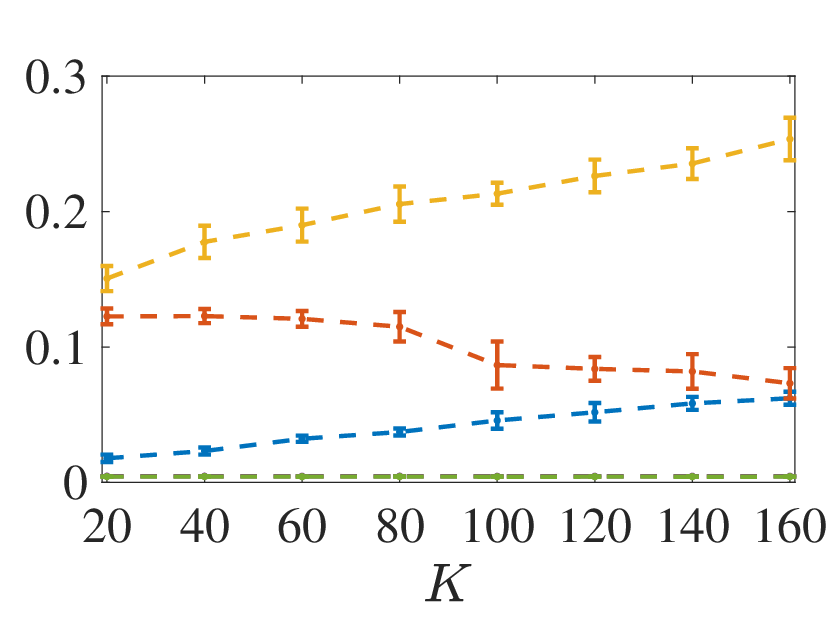}
        \caption{Cross-joint-ISI}
    \end{subfigure}
    \begin{subfigure}[b]{0.24\textwidth}
        \centering
        \includegraphics[width=\textwidth]{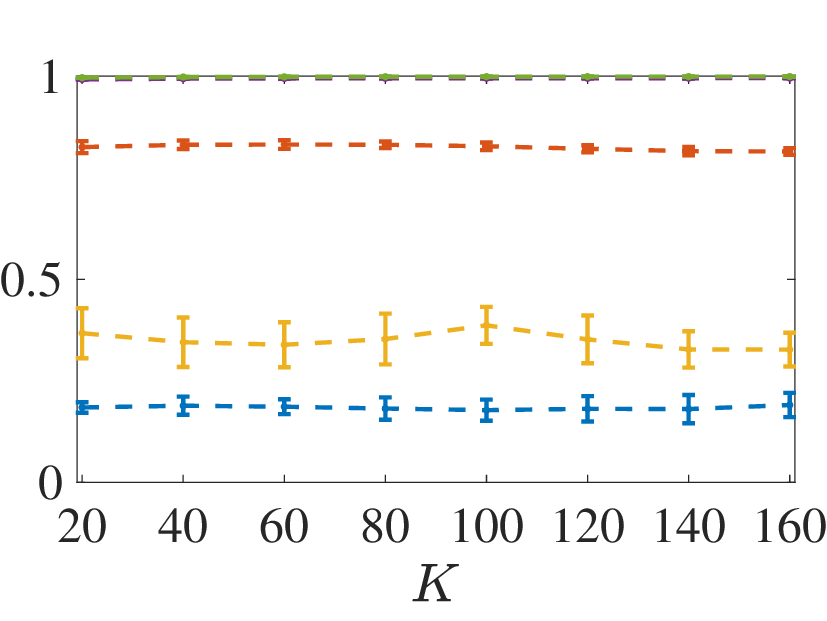}
        \caption{Similarity factor}
    \end{subfigure}
    \begin{subfigure}[b]{0.24\textwidth}
        \centering
        \includegraphics[width=\textwidth]{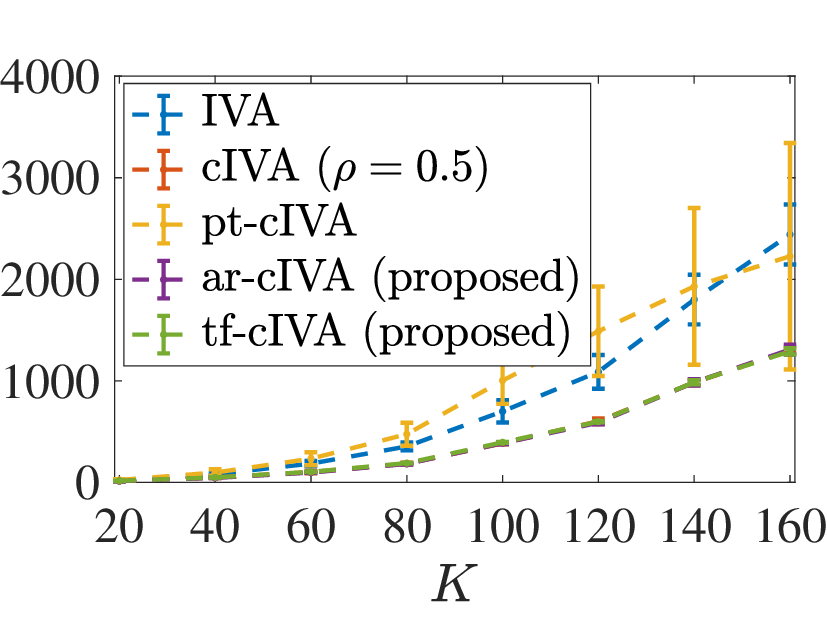}
        \caption{Runtime (seconds)}
    \end{subfigure}
    \caption{Comparison of five different IVA algorithms for the simulated fMRI-like data with $M = N = 20$ and $V=58515$. The (a) joint-ISI, (b) cross-joint-ISI, (c) similarity factor, and (d) runtime are shown as functions of the number of subjects $K$. The error bars represent one standard deviation calculated over $20$ runs. For each value of $K$, the same mixing matrix is used to generate the simulated fMRI-like data across $20$ runs. All algorithms use the same initialization for each run. Note that in all four plots, the purple line and the green line almost overlap and in plot (d), the red line also overlaps these two.}
    \label{fig:hybrid_result}
\end{figure*}

\noindent \textbf{Extraction of reference signals.}
We use reference signals extracted by NeuroMark, i.e., the Neuromark\_fMRI\_1.0 template\ \cite{du2020neuromark}, which includes $20$ fMRI networks and is divided into seven functional domains based on their anatomical and functional properties: the subcortical (SC), auditory (AUD), sensorimotor (MOT), visual (VIS), cognitive control (CC), default mode (DMN) and cerebellar (CB) domains.\footnote{The original template contains a total of $53$ references. In the hybrid simulation experiment with varying numbers of subjects, to reduce the runtime, we only use a subset of $20$ references (with $2$ references from AU and $3$ references from each of the other $6$ functional domains).} 
For convenience, we denote the set of $N=20$ reference signals by $\{ \bm r_n \}_{n=1}^N$, each contains $V=58515$ samples $\bm r_n = [r_n(1),r_n(2),\ldots,r_n(V)]^\topnew$. In addition, each reference signal is normalized to zero mean and unit variance. Finally, we note that there is a certain level of dependency among the reference signals, i.e., they are not absolutely independent.

\noindent \textbf{Hybrid source generation.}
Given the reference signals, we generate observations of SCVs $\{ \bm S_n \}_{n=1}^N \subset \R^{K \times V}$ for $K$ subjects as follows. First, we define a $NK$---dimensional random vector $\bm z$ following multivariate Gaussian distribution with zero mean and covariance matrix 
\begin{align*}
    \bm \Sigma_{\bm z} = \bigl(\mu_0 \bm 1_N \bm 1_N^\topnew + (\mu_1-\mu_0) \bm I_{N}\bigr) \otimes \bm 1_K \bm 1_K^\topnew + (1-\mu_1) \bm I_{NK} ,
\end{align*}
where $0 \leq \mu_0 \leq \mu_1 \leq 1$.
Second, we generate $V$ samples of $\bm z$ and partitioning the data matrix into $N$ submatrices of dimension $K \times V$, i.e., $\bm Z = [\bm Z_1^\topnew, \ldots, \bm Z_N^\topnew]^\topnew$.
Third, the $n$th source data matrix is formed by
\begin{align} \label{equ:s_hybrid}
    \bm S_n = \sqrt{1-\varphi_n^2} \bm 1_K \bm r_n^\topnew + \varphi_n \bm Z_n \quad \in \R^{K \times V},
\end{align}
where $\varphi_n \in [0,1]$ controls how close the $n$th source is to the reference $\bm r_n$. Figure~\ref{fig:hybrid_data}(c) depicts the SCV covariance matrices of the simulated fMRI-like data with various values of $\bm \varphi$ across source components. We emphasize that as $\varphi_n$ varies in the range $[0.3,0.9]$, the correlation between the source and the reference signal also varies in the range $[0.31,0.81]$.

\begin{figure*}[t]
    \centering
    \begin{subfigure}[b]{0.3\textwidth}
        \centering
        \includegraphics[width=\textwidth]{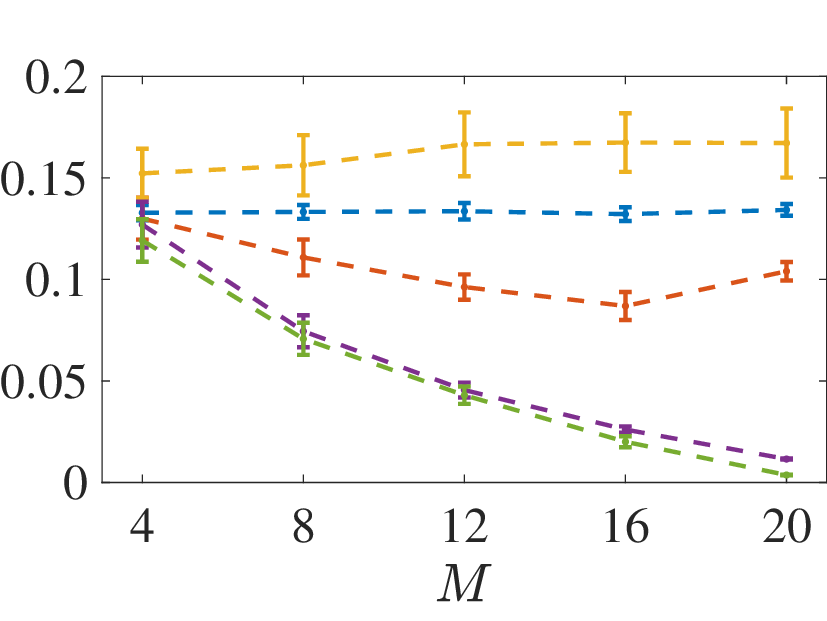}
        \caption{Joint-ISI}
    \end{subfigure}
    \quad
    \begin{subfigure}[b]{0.3\textwidth}
        \centering
        \includegraphics[width=\textwidth]{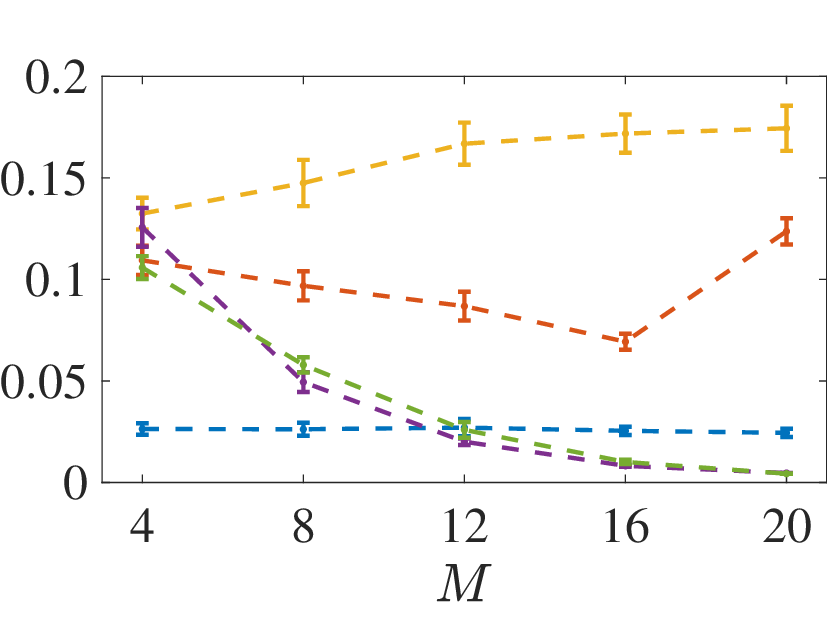}
        \caption{Cross-joint-ISI}
    \end{subfigure}
    \quad
    \begin{subfigure}[b]{0.3\textwidth}
        \centering
        \includegraphics[width=\textwidth]{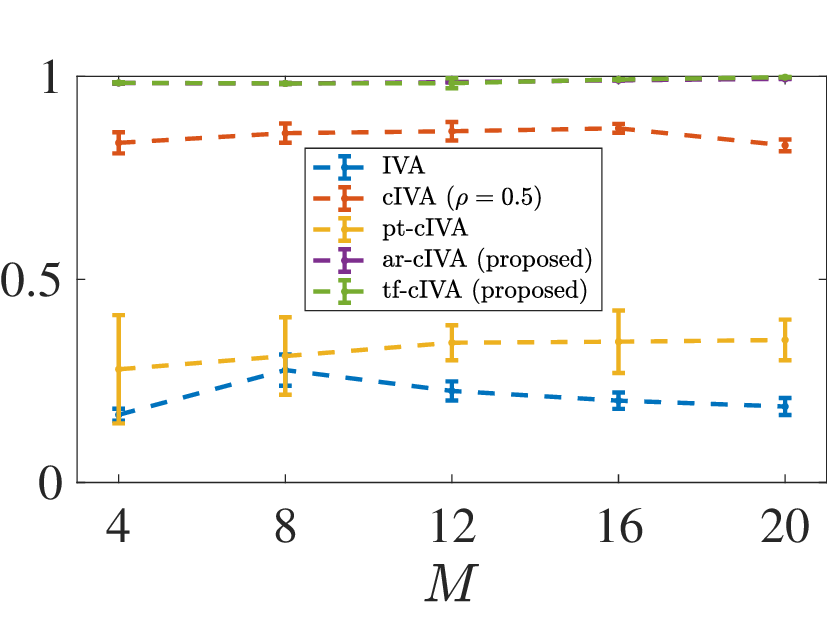}
        \caption{Similarity factor}
    \end{subfigure}
    \caption{Comparison of five different IVA algorithms for the simulated fMRI-like data with $N=20$, $K=40$, and $V=58515$ as the number of reference signals $M$ increases. The error bars represent one standard deviation calculated over $20$ runs. For each value of $R$, the mixing matrix remains the same across $20$ runs, while the hybrid sources are generated independently for each run. All algorithms use the same initialization for each run. Note that in plot (c), the purple line and the green line almost overlap.}
    \label{fig:hybrid_result_refs}
\end{figure*}

\noindent \textbf{Results.}
We evaluate the performance of the five aforementioned IVA algorithms with regard to the changes in (i) the number of subjects $K$ and (ii) the number of reference signals. In the first experiment, the number of subjects $K$ is varied while the numbers of samples $V$, source components $N$, and reference signals $M$ are fixed. As can be seen from Fig.~\ref{fig:hybrid_result}, our two proposed algorithms, tf-cIVA (dashed green line) and ar-cIVA (dashed purple line), significantly outperform other algorithms in terms of joint-ISI, cross-joint-ISI, and similarity factor. 
As the number of subjects $K$ increases, the cross-joint-ISI of the unconstrained IVA (the blue dashed line in Fig.~\ref{fig:hybrid_result}-b) decreases, indicating that this method becomes less reproducible in large-scale settings. This degradation is also noted in \cite{bhinge2019extraction} as the curse of dimensionality in IVA. The same phenomenon is also observed for pt-cIVA (yellow dashed line), which verifies our earlier discussion on the conservative nature of this method in selecting thresholds.
Interestingly, almost independent of the number of subjects, tf-cIVA and ar-cIVA yield consistently excellent performance. This is highlighted by the fact that our algorithms exploit both HOS (by effectively imposing constraints with reference signals) and SOS (via the Gaussian source model).
In terms of runtime, Fig.~\ref{fig:hybrid_result}-d shows that ar-cIVA and acIVA-fixed are the fastest algorithms while IVA is the slowest.

In the second experiment, we vary the number of reference signals $M$ while fixing the number of components $N$, the number of subjects $K$, and the number of samples $V$. The performance of the five aforementioned algorithms is shown in Fig.~\ref{fig:hybrid_result_refs}. We observe the effect of increasing the number of references in terms of joint-ISI and cross-joint-ISI: ar-cIVA (purple dashed line) and tf-cIVA (green dashed line) yield better separation results as $M$ increases while the performance of unconstrained IVA remains unchanged. The joint-ISI and cross-joint-ISI of the fixed-threshold scheme, cIVA with $\rho=0.5$ (red dashed line), also decreases as $M$ increases from $4$ to $16$ references. However, when the number of references equals the number of components, we observe a slight decrease in the performance of cIVA. This is because the correlation between the source and the reference signal varies in the range $[0.31,0.81]$ across components. Thus, for some components, the threshold $\rho=0.5$ cannot be satisfied. Finally, the pt-cIVA algorithm performs worst in terms of joint-ISI and cross-joint-ISI. Nonetheless, the similarity factor of pt-cIVA is higher than that of unconstrained IVA, indicating the adaptive rule in (\ref{equ:rho_pt}) is working to a certain degree but is not truly effective.

\section{Multi-Subject fMRI Data Analysis}
\label{sec:fMRI}

This section evaluates the performance of the different algorithms on real fMRI data. Our goal is to demonstrate that the proposed methods offer better model matches and more interpretable results. 

\noindent \textbf{Data acquisition and preprocessing.}
We use the resting state fMRI data set from the bipolar-schizophrenia network on intermediate phenotypes (B-SNIP) \cite{tamminga2013clinical,tamminga2014bipolar}. Identical diagnostic and recruitment approaches were applied to all recruited subjects at multiple sites (Baltimore, Chicago, Dallas, Detroit, and Hartford). In particular, in this study, we employ the data collected from the Baltimore site and select $K = 98$ subjects: 49 healthy controls (HCs) and 49 randomly selected schizophrenia patients (SZs). A single 5-minute run was captured for each subject. The individuals involved in the study were instructed to maintain an open-eyed state, concentrate on a crosshair presented on a display screen, and remain still throughout the scanning process. Moreover, a custom-built head-coil cushion was used to restrict head movements. Alertness during the scan was confirmed immediately afterward, and the procedure was repeated if needed. These instructions helped reduce head motion and prevented subjects from falling asleep. The fMRI data were captured by a 3-Tesla Siemens Triotim scanner with $\text{TE} = 30\text{ ms}$, $\text{TR}= 2.21\text{ s}$, $\text{flip angle} = 70^\circ$, $\text{acquisition matrix} = 64 \times 64\text{mm}$, and $\text{voxel size} =3.4 \times 3.4 \times 3 \text{ mm}^3$. For each subject, 134 time points were obtained. We removed the first 3 time points to address the T-1 effect and each subject’s image data was preprocessed including motion correction and slice-time correction. The corrected data were warped into the standard Montreal Neurological Institute (MNI) space through an echo-planar imaging template and then were resampled to  $3 \times 3 \times 3 \text{ mm}^3$ isotropic voxels.  The resampled fMRI data were further smoothed using a Gaussian kernel with a full width at half maximum (FWHM) equal to $6\text{ mm}$. In addition, in order to remove non-brain voxels and flatten the data, each subject image was masked, yielding an observation vector of $V = 58515$ voxels for each of the $T = 131$ time points.

\noindent \textbf{Results.}
We evaluate the performance of the five aforementioned IVA algorithms when applied to real fMRI data. For this purpose, as in the experiments developed for simulated fMRI-like data, we use the functional templates extracted by Neuromark, specifically the neuromark\_fMRI\_1.0 template\ \cite{du2020neuromark}, which is composed of 53 resting-state networks (RSNs) from seven different functional domains: SC (5 RSNs), AUD (2 RSNs), MOT (9 RSNs), VIS (9 RSNs), CC (17 RSNs), DMN (7 RSNs) and CB (4 RSNs). Each of these RSNs is employed as a reference signal by the IVA algorithms. In addition, since fMRI data can also contain signals not of interest such as motion-related signals, scanner-related signals, or noise due to magnetic resonance acquisition, among others, we use a larger number of components than reference signals ($N > 53$) so that we can capture those signals not of interest in the free components not related to a specific functional template. In our experiments, we noted $N =60$ as a good number that balances the trade-off between the model complexity and flexibility.

\begin{figure*}[t]
    \centering
    \begin{subfigure}[b]{0.32\textwidth}
        \centering
        \includegraphics[width=\textwidth]{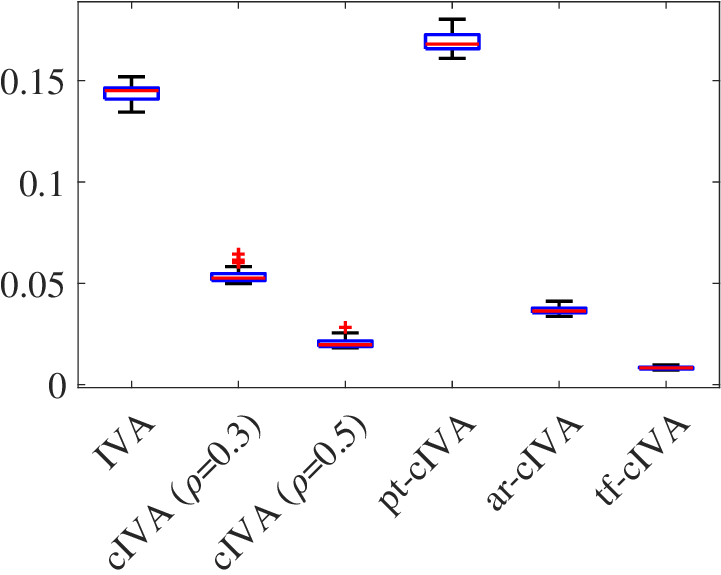}
        \caption{Cross-joint-ISI}
    \end{subfigure}
    \quad
    \begin{subfigure}[b]{0.32\textwidth}
        \centering
        \includegraphics[width=\textwidth]{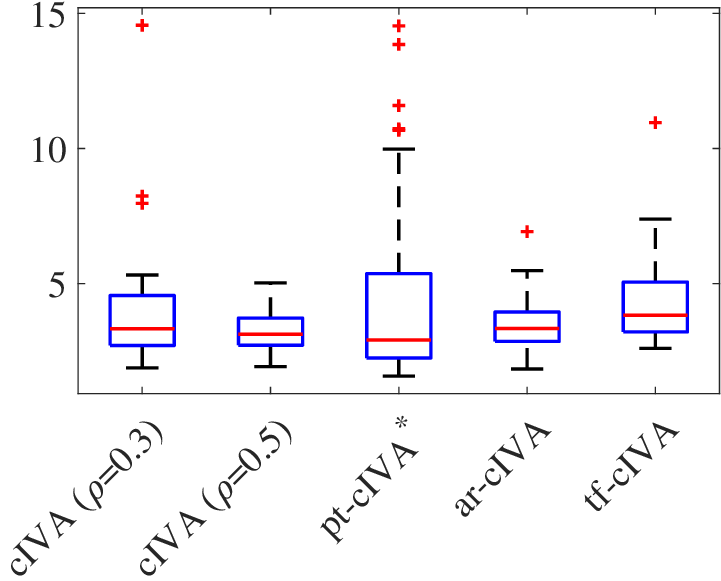}
        \caption{Power ratio}
    \end{subfigure}
    \quad
    \begin{subfigure}[b]{0.25\textwidth}
        \centering
        \includegraphics[width=\textwidth]{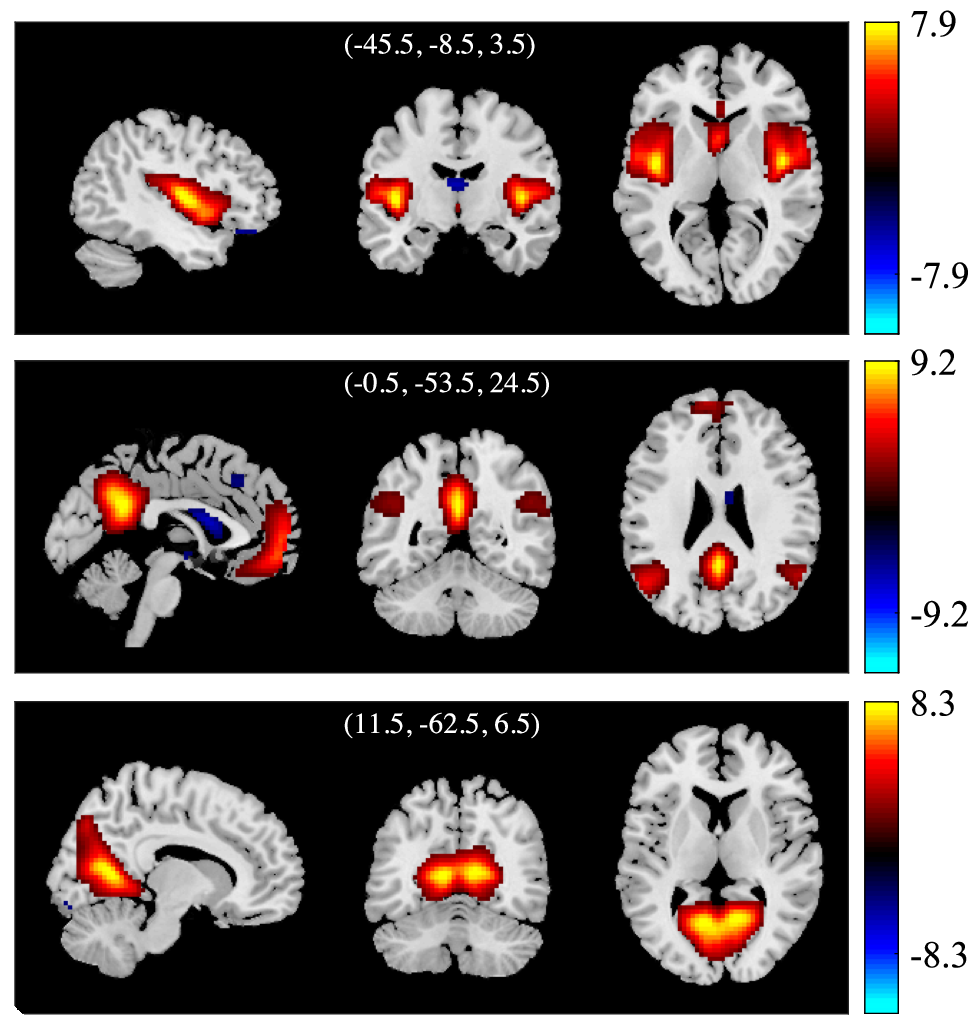}
        \caption{Spatial maps}
    \end{subfigure}
    \caption{Comparison of different IVA algorithms for real fMRI data with $N=60$, $K=98$, and $V=58515$. Plot (a) shows the cross-joint-ISI values for 50 independent runs for each algorithm. Plot (b) shows the power ratio for the most consistent run of the constrained algorithms. Note that in the case of pt-cIVA three outliers (with values of $22.59$, $26.92$, and $29.57$) were removed for a clearer comparison. Plot (c) shows the average spatial maps across the $K=98$ subjects of three different RSNs: AUD, DMN, VIS (top to bottom). The coordinates (mm) of the peak activity are shown at the top of each spatial map.}
    \label{fig:real_results}
\end{figure*}

The obtained results by the different IVA algorithms are shown in Fig.~\ref{fig:real_results}. The cross-joint-ISI values for 50 independent runs are depicted in Fig.~\ref{fig:real_results}-a. Unconstrained IVA and pt-cIVA present higher values than the rest of the algorithms, achieving less consistent results across all the runs. It can also be seen that the fixed-threshold scheme, cIVA with $\rho = 0.5$, is the second-best algorithm in terms of cross-joint-ISI. However, if the threshold value is modified, cIVA with $\rho = 0.3$, its performance decreases, making it clear that the correct selection of a threshold value is of paramount importance for the performance of these algorithms. On the other hand, we can also observe that the proposed adaptive-reverse scheme, ar-cIVA, also offers low cross-joint-ISI values with small variance across runs and significantly outperforms the previously proposed adaptive cIVA algorithm (pt-cIVA). Finally, we can see that the proposed algorithm tf-cIVA outperforms the rest of the algorithms, achieving the most consistent results with the lowest cross-joint-ISI values. The spatial maps of three different RSNs (AUD, DMN and VIS) obtained by the most consistent run of this algorithm (tf-cIVA) are shown in Fig.~\ref{fig:real_results}-c as an example.

Another useful measure to quantify the quality of estimation of the fMRI components is the power spectra of RSN time courses and the power ratio between low-frequency ($<0.1\,\text{Hz}$) and high-frequency ($>0.15\,\text{Hz}$) bands. Considering the frequencies of neural-activity related BOLD signals are generally below 0.15 Hz, low power ratio values are typically associated with cardiac and respiratory noise, while high power ratio values mostly indicate BOLD activity \cite{allen2011baseline}. The power ratio for the most consistent run of each algorithm is depicted in Fig.~\ref{fig:real_results}-b. It is important to note that in the case of unconstrained IVA, only 26 out of the 60 estimated components are selected as meaningful after inspecting their spatial maps and power spectra values, hence its power ratio results are omitted in Fig.~\ref{fig:real_results}-b. In addition, for the sake of a clearer comparison of the constrained algorithms, three outliers from pt-cIVA (with power ratio values of $22.59$, $26.92$, and $29.57$) were removed. For the other algorithms, the 53 estimated components related to the reference signals are taken into account for their assessment. As we can observe, the proposed algorithm tf-cIVA achieves the highest median power ratio value (red line within each boxplot) followed by ar-cIVA. Since high power ratio values are associated with higher BOLD activity, the higher values obtained by tf-cIVA indicate better performance in the estimation of the components.

\begin{figure*}[t]
    \centering
    \begin{subfigure}[b]{0.31\textwidth}
        \centering
        \includegraphics[width=\textwidth]{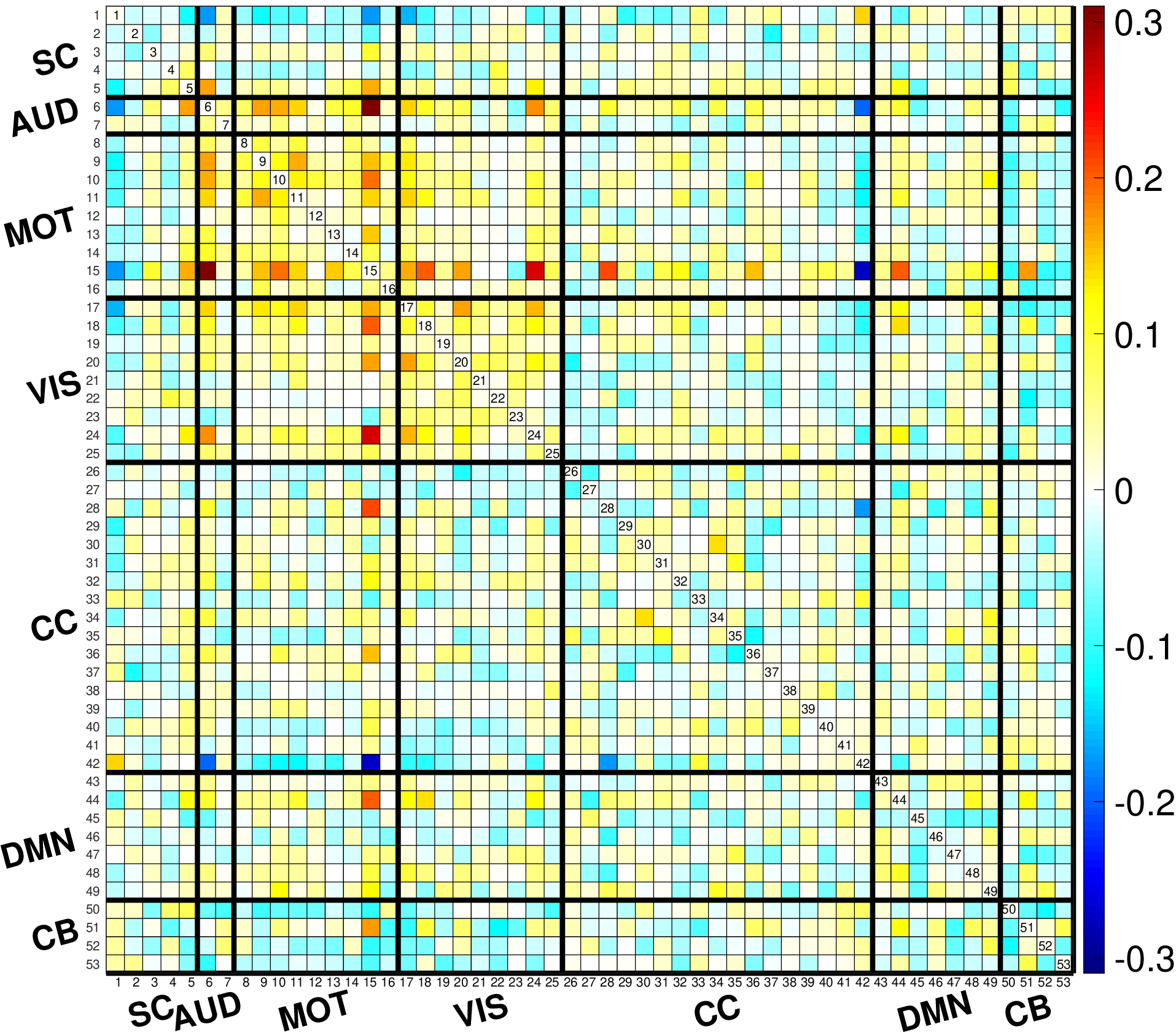}
        \caption{Average FNC cIVA ($\rho=0.3$)}
    \end{subfigure}
    \quad
    \begin{subfigure}[b]{0.31\textwidth}
        \centering
        \includegraphics[width=\textwidth]{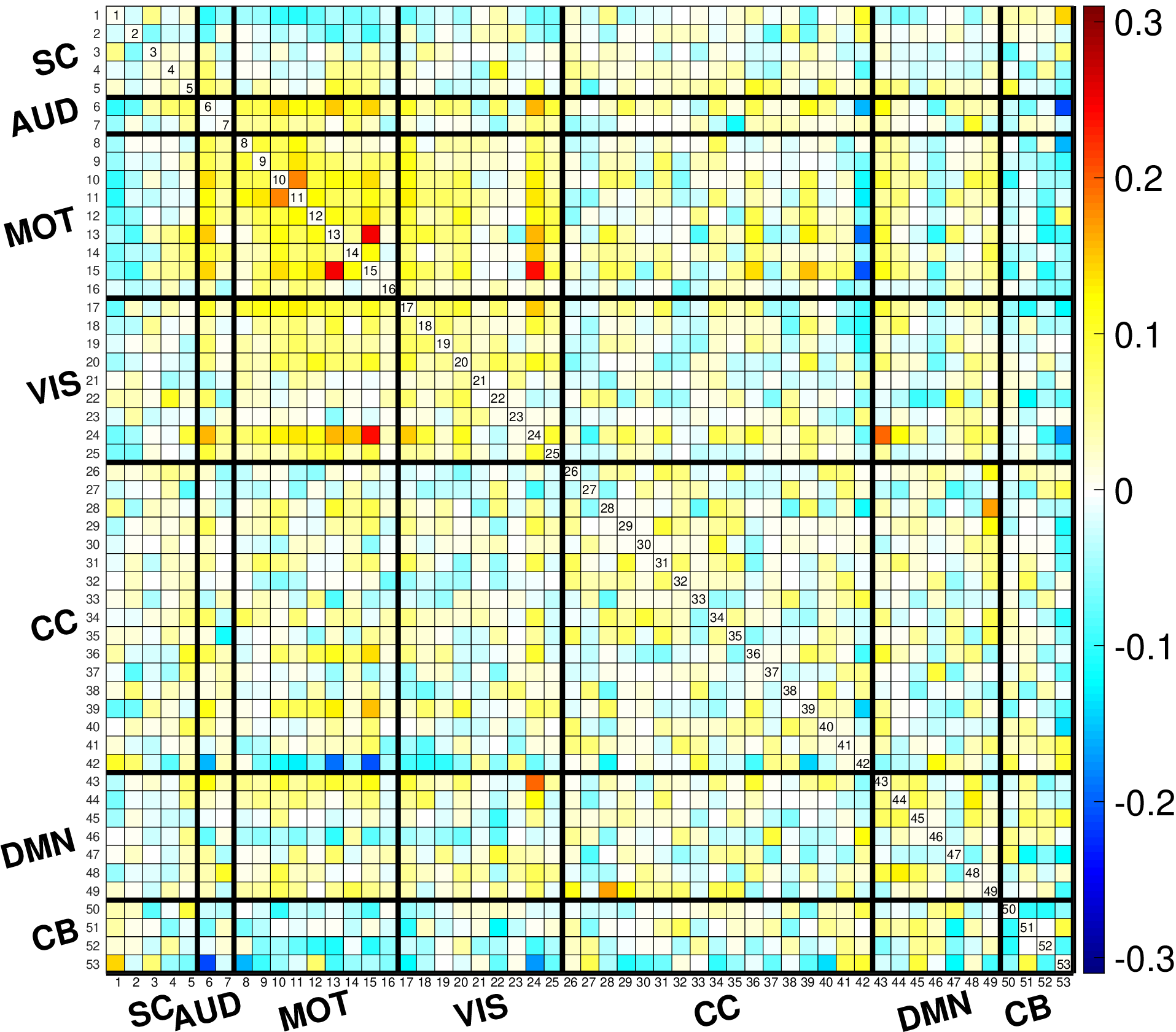}
        \caption{Average FNC ar-cIVA}
    \end{subfigure}
    \quad
    \begin{subfigure}[b]{0.31\textwidth}
        \centering
        \includegraphics[width=\textwidth]{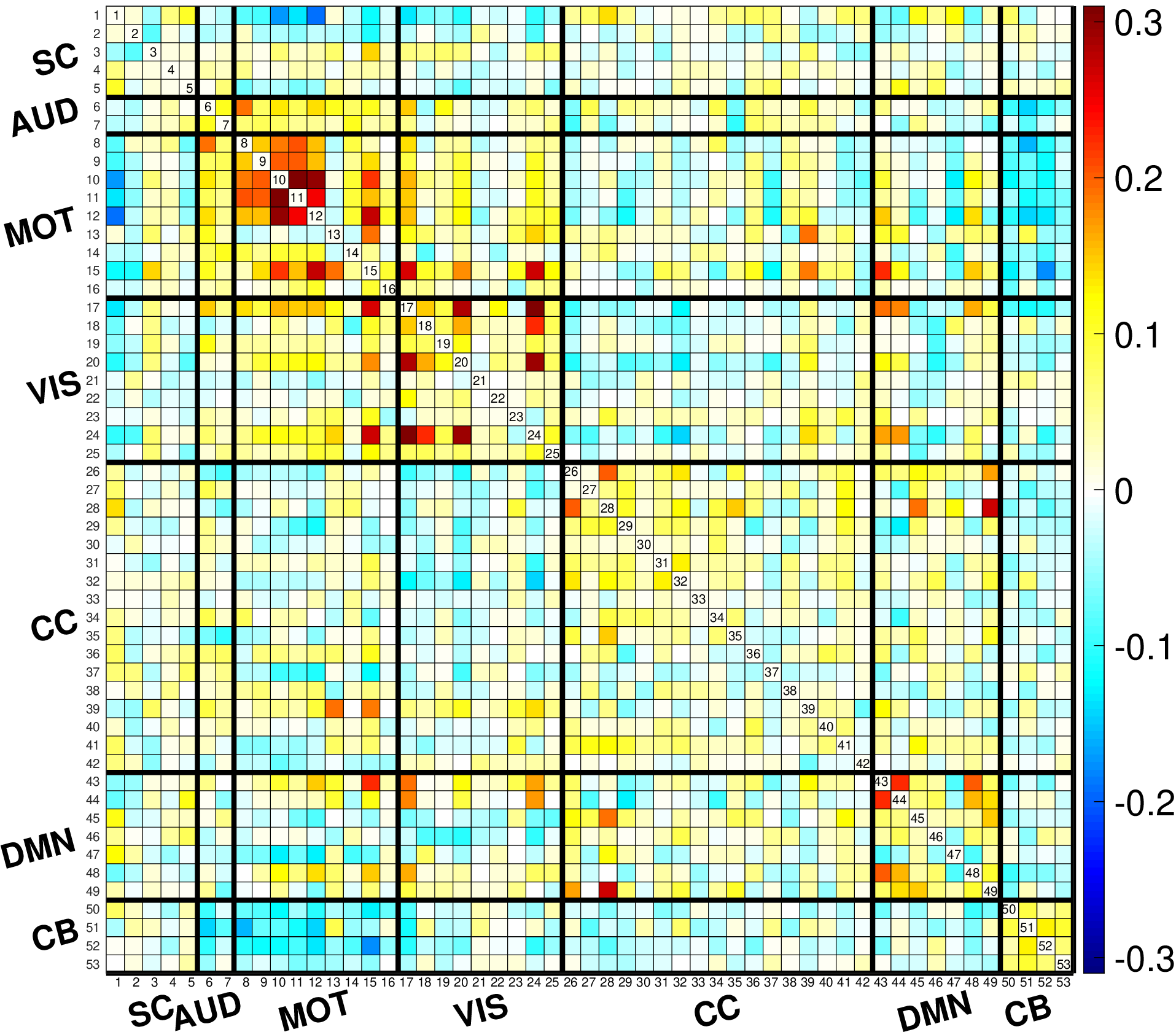}
        \caption{Average FNC tf-IVA}
    \end{subfigure}

    \caption{Aggregated FNC matrix for the most consistent run. Pairwise Pearson correlation between RSNs time courses are first Fisher z-transformed and averaged across all subjects, then inverse z-transformed for display. The 53 components associated with the reference signals are considered for the analysis of the FNC matrix.}
    \label{fig:real_FNCs}
\end{figure*}

\begin{figure*}[t]
    \centering
    \begin{subfigure}[b]{0.31\textwidth}
        \centering
        \includegraphics[width=\textwidth]{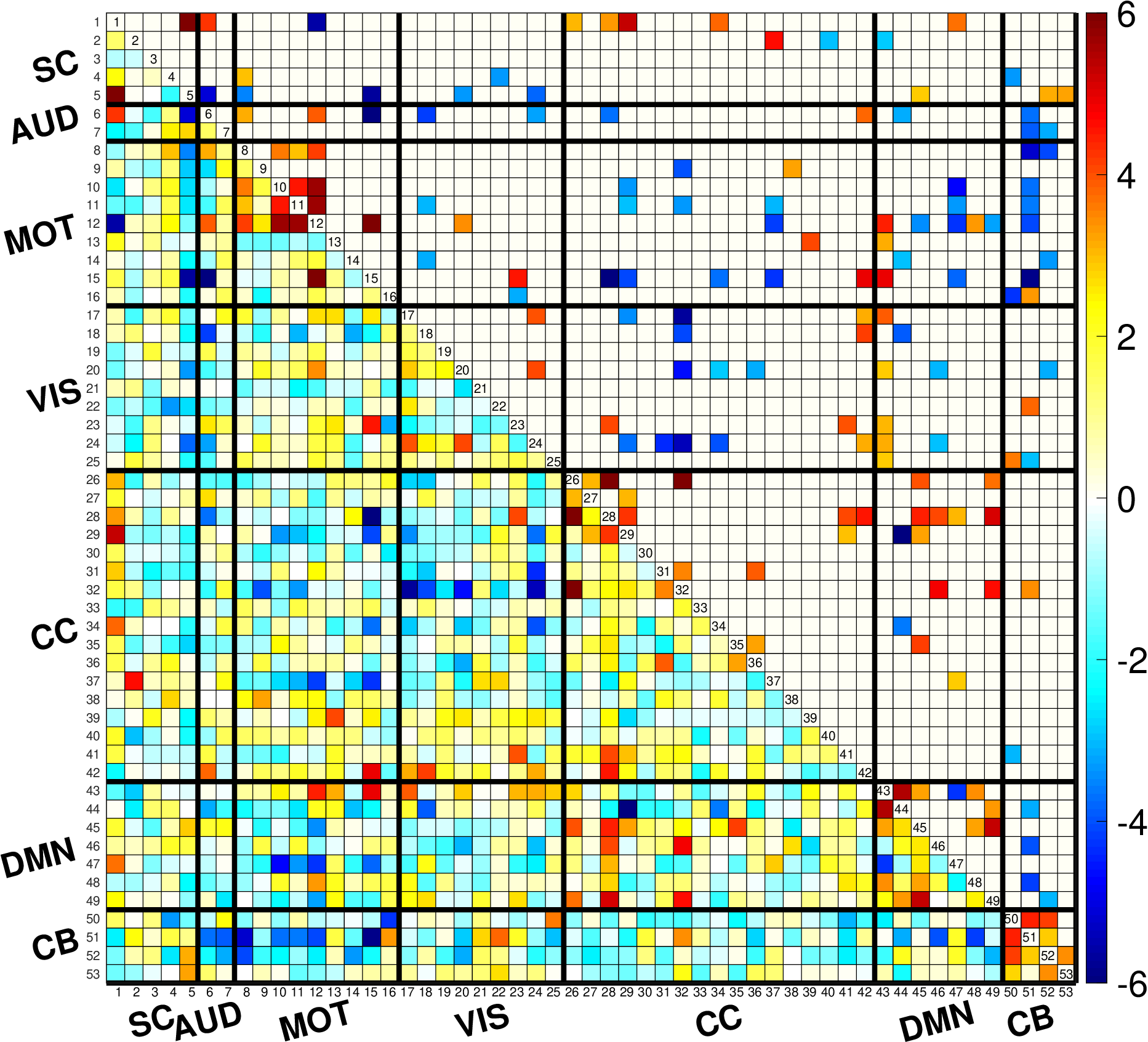}
        \caption{T-values cIVA ($\rho=0.3$)}
    \end{subfigure}
    \quad
    \begin{subfigure}[b]{0.31\textwidth}
        \centering
        \includegraphics[width=\textwidth]{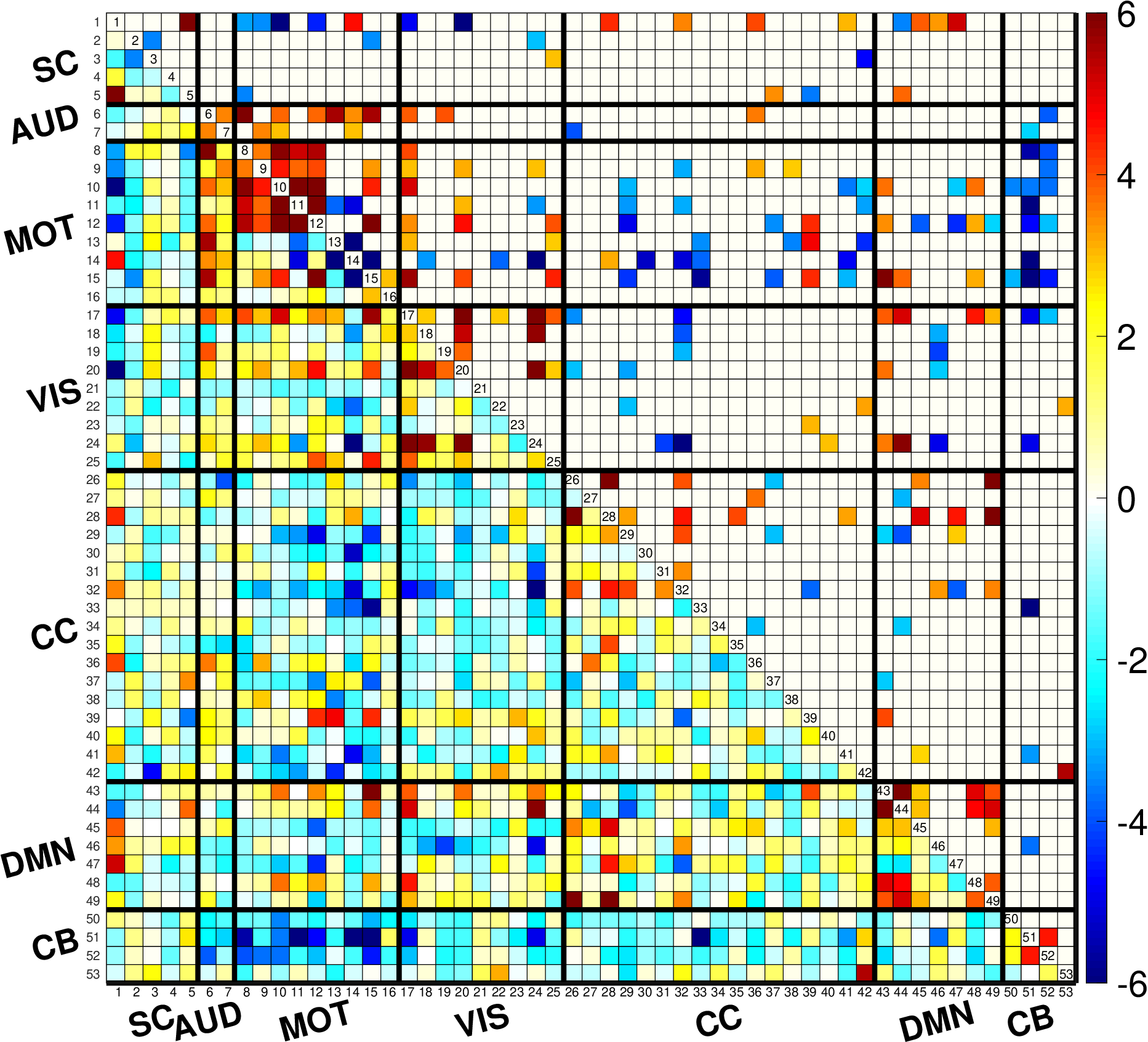}
        \caption{T-values pt-cIVA}
    \end{subfigure}
    \quad
    \begin{subfigure}[b]{0.31\textwidth}
        \centering
        \includegraphics[width=\textwidth]{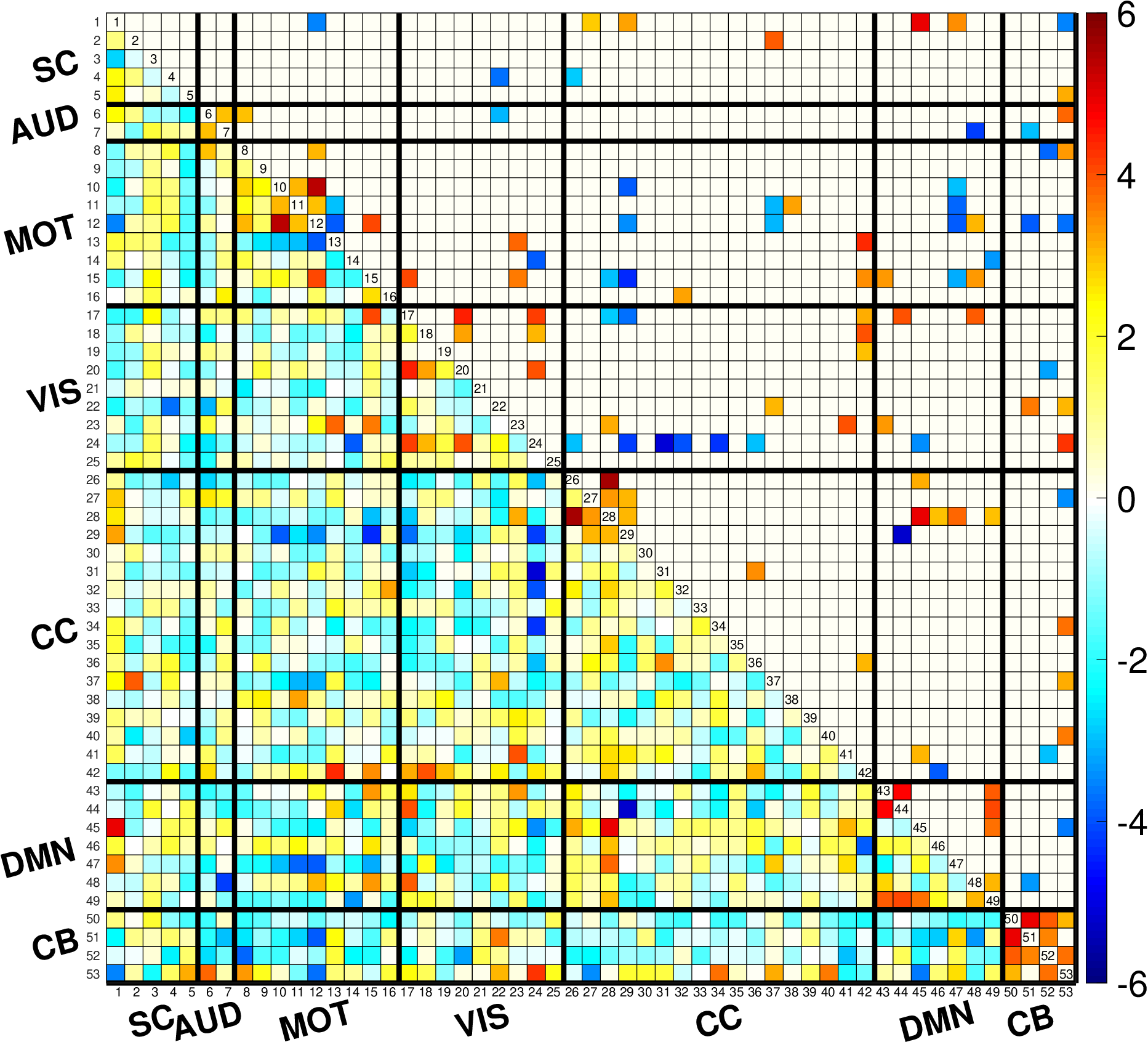}
        \caption{T-values ar-IVA}
    \end{subfigure}

    \caption{T-value maps showing the differences in the FNCs obtained by paired t-tests for tf-cIVA vs. cIVA ($\rho=0.3$), tf-cIVA vs. pt-cIVA, and tf-cIVA vs. ar-cIVA. The lower diagonal of the maps shows the T-values before the FDR correction, while the upper diagonal shows the T-values after passing the FDR correction (p-value $< 0.05$). Positive T-values indicate tf-cIVA shows a higher connectivity value than the compared algorithm. The 53 components associated with the reference signals are considered for the paired t-test.}
    \label{fig:real_t-paired}
\end{figure*}

For a more detailed analysis of the algorithms, we also evaluate the correlations between components' time courses, i.e., the functional network connectivity (FNC) maps. The patterns revealed in FNC are consistent with known functional network relationships, for example, we note the anticorrelation between the DMN components and sensory-related networks (e.g., MOT and VIS) \cite{allen2011baseline}. The aggregated FNC matrices for the most consistent run for the three constrained algorithms with the highest median power ratio are shown in Fig.~\ref{fig:real_FNCs}. The FNC obtained by the proposed algorithm tf-cIVA shows higher contrast than the rest of the algorithms. In particular, we can see a positive correlation between functional domains such as sensorimotor and visual, and a negative correlation between DMN and sensory-related networks. On the other hand, the FNC matrices obtained by the other algorithms do not show such a clear pattern, where the correlation value is, for most of the RSNs, closer to zero. In addition, we also conduct a statistical analysis to compare the performance of the algorithms. For this purpose, a paired t-test of the FNCs matrices is performed, and the results obtained by tf-cIVA are compared with those obtained by the fixed-threshold algorithm with the highest power ratio, cIVA ($\rho=0.3$), and with the two other adaptive algorithms, i.e., pt-cIVA and ar-cIVA. The resulting T-value maps are shown in Fig.~\ref{fig:real_t-paired}, where the upper diagonal presents the networks with significant connectivity differences between the compared algorithms after the false discovery rate (FDR) correction of the p-values ($< 0.05$) \cite{benjamini2005false}. It can be seen that tf-cIVA exhibits higher connectivity values within the functional domains such as AUD, VIS, MOT, DMN, or CB. To quantify this performance, we compute the percentage of RSNs showing a significant difference within each functional domain where tf-cIVA achieves higher connectivity values. When compared with cIVA ($\rho=0.3$), tf-cIVA achieves higher connectivity values in $96.88\,\%$ of the RSNs, when compared with pt-cIVA the percentage achieved is $83.33\,\%$ and $93.10\,\%$ for ar-cIVA. Hence, the obtained FNCs and T-maps suggest that tf-cIVA provides a better model match to fMRI data and therefore more interpretable results \cite{adali2022reproducibility}.

\begin{figure*}[t]
    \centering
    \flushleft
    \begin{subfigure}[b]{0.45\textwidth}
        \centering
        \includegraphics[width=\textwidth]{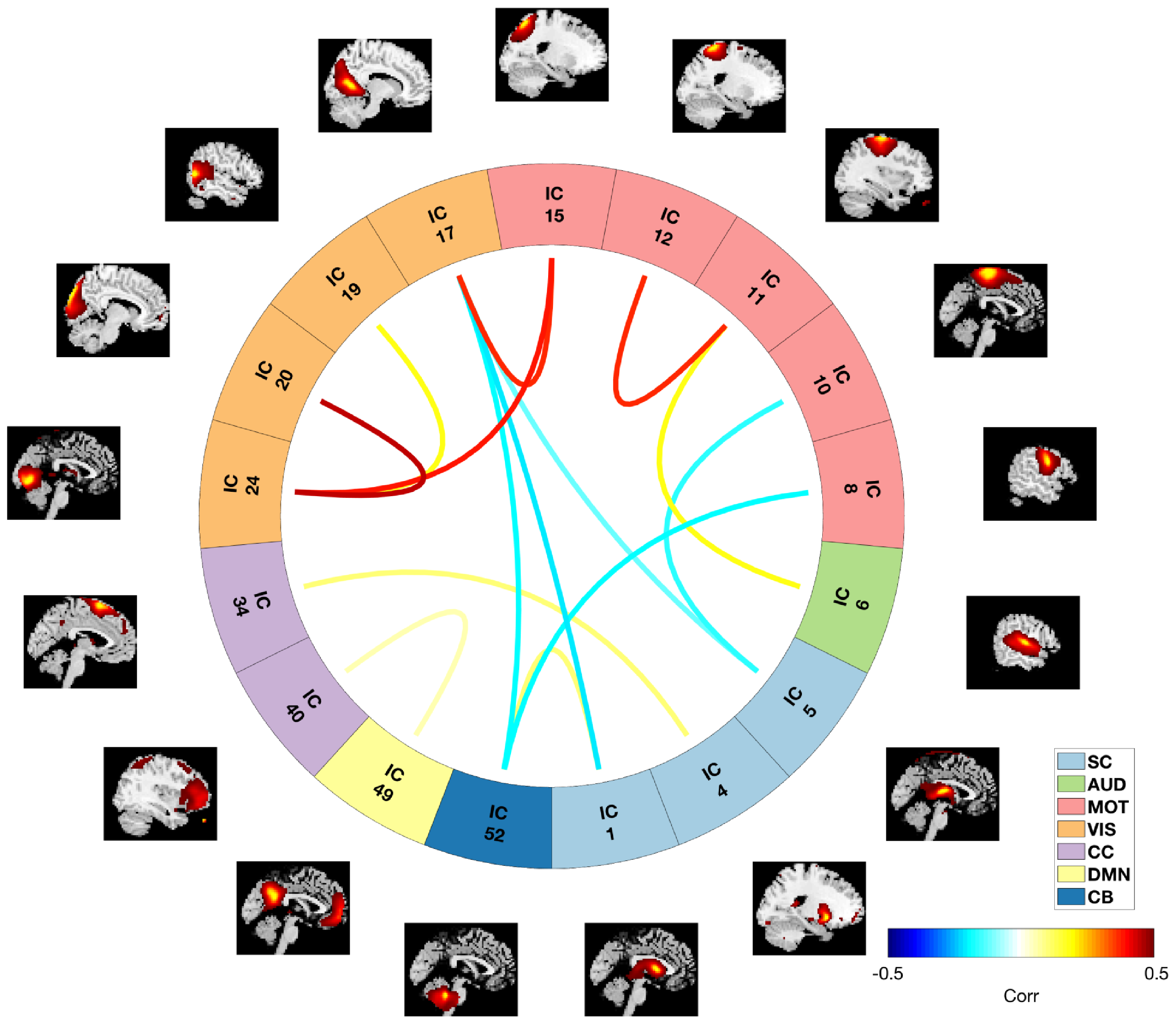}
        \caption{Average FNC HC}
    \end{subfigure}
    \quad
    \begin{subfigure}[b]{0.45\textwidth}
        \centering
        \includegraphics[width=\textwidth]{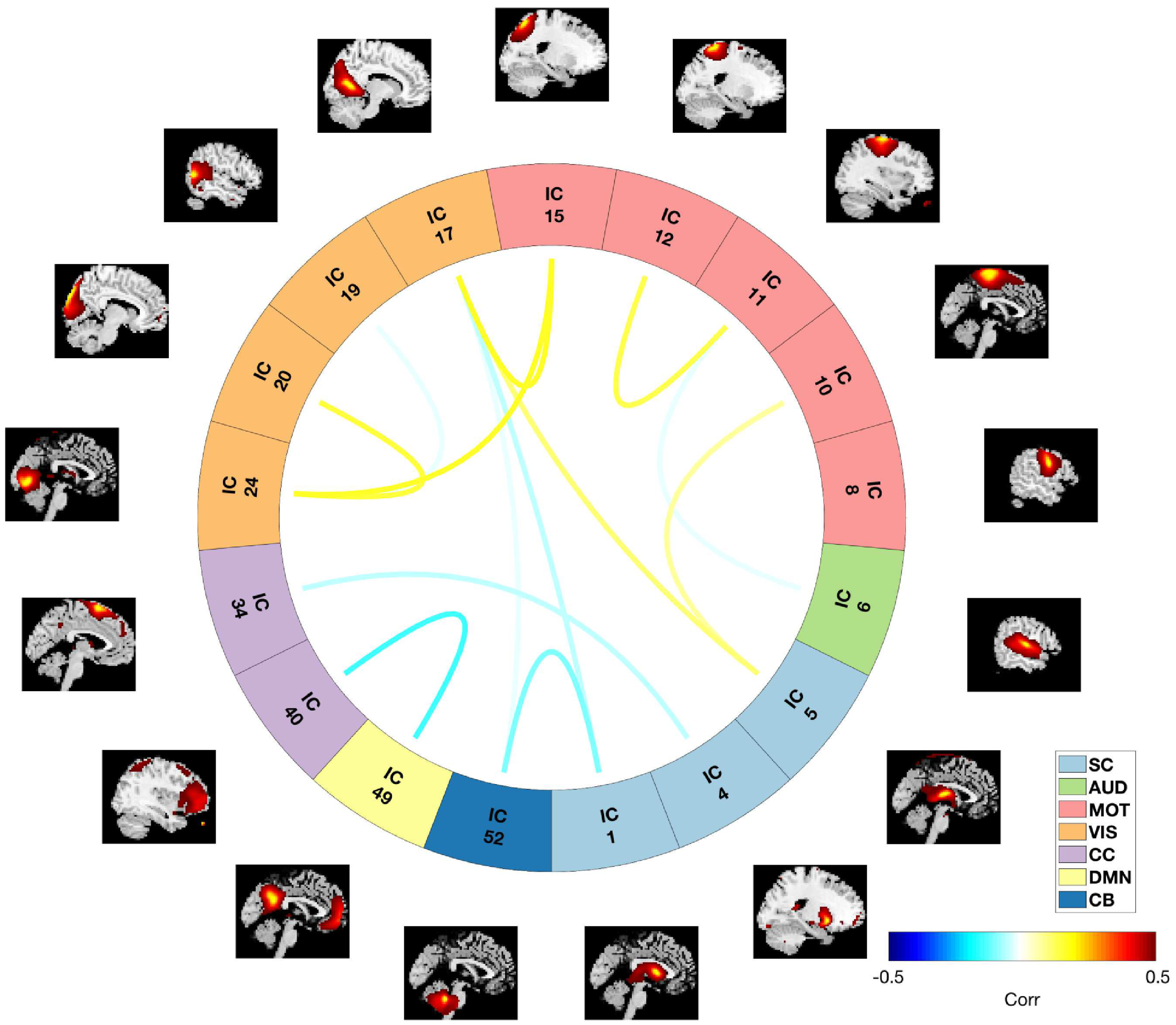}
        \caption{Average FNC SZ}
    \end{subfigure}

    \caption{Average FNC strength obtained by tf-cIVA across subjects for HC and SZ groups. A two-sample t-test is applied and only the RSNs with significant differences between groups after FDR correction are shown. For each RSN, the averaged values in SZ and HC groups are presented by the connecting lines. The outer circle depicts the ICs indices and their corresponding average spatial maps. The 53 components associated with the reference signals are considered for the two-sample t-test.}
    \label{fig:real_connectograms}
\end{figure*}

Furthermore, we also analyze the group differences of the FNCs between HC and SZ. To this end, a two-sample t-test is applied to the results obtained by the algorithms. After FDR correction of the p-values, tf-cIVA is the only algorithm showing more than one significant connectivity difference between groups, i.e., cIVA ($\rho=0.5$) and ar-cIVA do not show any significant difference, while cIVA ($\rho=0.3$) and pt-cIVA show significant differences only between two RSNs. Fig.~\ref{fig:real_connectograms} demonstrates the mean FNC of the networks that have a significant difference between groups after FDR correction of the p-values. The results show that SZ patients present weaker connectivity than HC within MOT and VIS domains. Also, the SZ group shows lower connectivity between domains such as MOT and AUD, and MOT and VIS. Some additional group differences can be observed between CB and MOT and VIS, and also between SC and MOT and VIS RSNs, consistent with recent studies \cite{du2021evidence,meng2022multimodel, du2020neuromark}. These group differences observed in tf-cIVA results increase our confidence that the proposed algorithm results in a better model match and performance in preserving subject variability.

\section{Discussion}
\label{sec:conc}

We proposed two novel approaches for constrained IVA that alleviate the need for pre-specified thresholds, thus significantly increasing their utility for fMRI data analysis. We demonstrated that these methods yield fully interpretable network estimates and can effectively capture HOS, even though they are implemented with a multivariate Gaussian model. The multivariate Gaussian implementation along with the use of an effective constraint framework hence enables achieving a desirable balance between performance and computational complexity.  An additional advantage of the constrained approach is that the permutation ambiguity of ICA/IVA is alleviated and post-analysis and sorting of components becomes a much easier task. 

While we have demonstrated here an application to a dataset with 98 subjects, significantly higher than what has been used with IVA approaches that take HOS into account, and higher than results with IVA-G, the method is scalable to the analysis of thousands of subjects. Hence, the proposed methods enable large-scale analyses, including the identification of homogeneous subgroups, studies of large-scale dynamics, and replicability. The new methods are also applicable to other joint BSS applications such as those in remote sensing and video analysis. 

\bibliographystyle{IEEEtran}
\bibliography{IEEEabrv,refs}

\end{document}